\documentclass[conference]{IEEEtran}
\IEEEoverridecommandlockouts
\usepackage{amsmath}
\usepackage{graphicx}%
\usepackage{epstopdf}
\usepackage{amsfonts}%
\usepackage{amssymb,mathrsfs}
\usepackage{algorithm}%
\usepackage{breakurl}
\usepackage{cite}
\usepackage{url}
\usepackage{flushend}
\usepackage{xcolor}
\usepackage{threeparttable}
\usepackage{enumerate}
\usepackage{verbatim}

\begin{document}
\title{Scenario-based Economic Dispatch with Uncertain Demand Response}
\author{Hao~Ming,~\IEEEmembership{Student Member,~IEEE,}
        Le~Xie,~\IEEEmembership{Senior Member,~IEEE,}
        Marco~Campi,~\IEEEmembership{Fellow,~IEEE,}\\
        Simone~Garatti,~\IEEEmembership{Member,~IEEE,}
        and~P. R.~Kumar,~\IEEEmembership{Fellow,~IEEE}% <-this % stops a space

%\thanks{This work is supported in part by NSF CyberSEES 1331863 and ECCS 1546682.}}
\thanks{H. Ming, L. Xie and P. R. Kumar are with the Department
of Electrical and Computer Engineering, Texas A\&M University, College Station,
TX, 77843-3128 USA e-mail: {mingming511}@tamu.edu, {le.xie}@tamu.edu.}% <-this % stops a space
\thanks{M. Campi is with the Department
of Information Engineering, University of Brescia, via Branze, 38, 25123 Brescia ITALY.}% <-this % stops a space
\thanks{S. Garatti is with the Dipartimento di Elettronica, Informazione e Bioingegneria, Politecnico
di Milano, ITALY.}% <-this % stops a space
\thanks{This work is supported in part by NSF Contract ECCS-1546682, NSF Science \& Technology Center Grant CCF-0939370, Electric Reliability Council of Texas (ERCOT), and the Power Systems Engineering Research Center (PSERC).}% <-this % stops a space
\thanks{Preferred address for correspondence: Le Xie, Dept. of ECE, Texas A\&M Univ., 3128 TAMU, College Station, Tx 77843-3128.}
}

%\IEEEauthorblockA{Department of Electrical and Computer Engineering, Texas A\&M University \\College Station, TX 77843-3128, USA\\
%Email: {mingming511}@tamu.edu, {Le.Xie}@tamu.edu}}
%Email: {default.email}@tamu.edu}}
\maketitle
% \footnote{This work is supported in part by NSF CyberSEES CCF-1331863.}

\begin{abstract}
%\boldmath

This paper introduces a new computational framework to account for uncertainties in day-ahead electricity market clearing process in the presence of demand response providers. A central challenge when dealing with many demand response providers is the uncertainty of its realization. In this paper, a new economic dispatch framework that is based on the recent theoretical development of the scenario approach is introduced. By removing samples from a finite uncertainty set, this approach improves dispatch performance while guaranteeing a quantifiable risk level with respect to the probability of violating the constraints. The theoretical bound on the level of risk is shown to be a function of the number of scenarios removed. This is appealing to the system operator for the following reasons: (1) the improvement of performance comes at the cost of a quantifiable level of violation probability in the constraints; (2) the violation upper bound does not depend on the probability distribution assumption of the uncertainty in demand response. Numerical simulations on (1) 3-bus and (2) IEEE 14-bus system (3) IEEE 118-bus system suggest that this approach could be a promising alternative in future electricity markets with multiple demand response providers.

%%%%%%%%%%%%%%%%%%%%%%%%%%%%%%%%%%%%%%%%%%%%%%%%%%%%%%%%%%%%%%%%%%%%%%%%%%%%%%%%%%%%%%%%%%%%%%%%%%%%%%%%%
% related MATLAB files:
% 1) ERCOT data: C:\Users\hao\Dropbox\Working File\Research\2_DR Benefit\3_DemandResponseRatio
% 2) Simulation: C:\Users\hao\Dropbox\Working File\Research\2_DR Benefit\2_Stochastic_DAM\V9.5
%%%%%%%%%%%%%%%%%%%%%%%%%%%%%%%%%%%%%%%%%%%%%%%%%%%%%%%%%%%%%%%%%%%%%%%%%%%%%%%%%%%%%%%%%%%%%%%%%%%%%%%%%

\end{abstract}

\begin{IEEEkeywords}
Demand response provider, scenario approach, stochastic economic dispatch
\end{IEEEkeywords}

\IEEEpeerreviewmaketitle

\section*{Nomenclature}

\begin{tabular}
{cl}
% {cp{2.5in}}
%      DR  & demand response\\
      ISO & independent system opeartor\\
      LSE & load serving entity\\
      DRP & demand response provider\\
      $P_{G,i}$ & power generation of a generator\\
      $\overline{P_{DR,j}}$ & maximum DR commitment of a DRP\\
      $P_{DR,j}$ & accepted DR commitment of a DRP\\
      $P_{base}$ & baseline for end-consumers\\
      $\pi_{s,j}$ & DRP's incentive price to end-consumers\\
      $\pi_{DR,j}$ & a DRP's offer price\\
      $\delta_{j}$ & DR ratio for DRP\\
      $P_{DR,j}^{real}$ & realized DR commitment\\
      $\xi$ & decision variable vector\\
      $P_{Lk}$ & total load on bus $k$\\
      $\alpha$ & friction factor for re-dispatch cost\\

\end{tabular}
%\end{table*}

\begin{tabular}
{cl}

      $\epsilon$ & constraint violation probability\\
      $\beta$ & confidence level of $\epsilon$\\
      $U^{s},U^{r}$ & uncertainty set for $\delta$ in stochastic/robust model\\
      $N_{k}$ & number of samples in the uncertainty set\\
      $p$ & number of samples removed\\
      $d$ & number of decision variables\\
\end{tabular}

\section{Introduction}
% Draft of Introduction
% 1) DR important and more & more
This paper explores a promising alternative for managing uncertainties in power system optimization. The context of this paper lies in the day-ahead scheduling of resources with many demand response providers. There has been an increasing need for demand response (DR) since the integration of renewables such as solar and wind requires more flexibility from the demand side. As an intermediate between the independent system operator (ISO) and end-consumers, demand response providers (DRPs) are permitted to collect demand response commitments from end-consumers through various means of incentives, and are treated as ``virtual generators" in most wholesale market clearing processes \cite{CAISO_DR,NYISO_DR,PJM_DR}.

% 2) DR unlike traditional generators in 1) uncertain 2) bottom up aggregation
However, unlike conventional generators whose outputs can be fully controlled, ``virtual generation" (demand reduction commitments) from DRPs is much less controllable \cite{Uncertainty_Mathieu1,Uncertainty_Mathieu2}. This uncertainty primarily comes from the fact of bottom-up aggregation from individual end-consumers, whose behavior is highly uncertain.

There has been a large body of literature investigating different optimization tools to manage the uncertainties in the economic dispatch problem. Stochastic approaches focus on maximizing expected social welfare \cite{Stochastic_Classic,Stochastic_Zhao,Stochastic_Dhillon}. However, this would require prior knowledge of the probability distribution of the underlying uncertainty. More recently, a number of papers investigate the robust optimization-based approach to unit commitment and economic dispatch \cite{Anupam_2014,Sun_Robust}. It provides a worst case optimal dispatch given an uncertainty set. However, the conservativeness of the robust optimization and the computational burden require substantial efforts to overcome.

As an algorithmic implementation of chance-constrained optimization, the scenario approach \cite{Campi_2009,Campi_2011} is introduced in this paper in the day-ahead market (DAM) clearing process. The key advantage of using this approach is that it allows the trade-off between robustness (constraint violation probability) and performance (objective function value) \cite{ProbN1_Vrakopoulou}.

The contributions of this paper are the following:
\begin{enumerate}[(1)]
\item It is the first paper that applies the scenario approach to solving the power system economic dispatch problem in the presence of uncertain demand response providers. Compared with traditional methods such as stochastic and robust optimization which only provide one solution, by sampling from the uncertainty set and gradually removing those scenarios, the scenario approach provides a curve of realization cost versus samples removed or constraint violation probability (Fig. 6(a)(b)).
\item Unlike traditional methods, the scenario approach can be used in the economic dispatch without any assumption on the distribution of the underlying uncertainty. Therefore, the scenario approach could potentially help avoid social welfare loss due to the lack of knowledge of the true underlying uncertainty distribution.
\item We investigate the impact of several different scenario removal algorithms on the violation of the categories of constraints (such as power adequacy, branch flow etc.). It sheds lights on system operators' choice of which subset of scenarios to be removed based on their priorities.
\end{enumerate}

This paper is organized as follows. Section II formulates a model to describe the uncertainty of DRPs in day-ahead market. Multiple economic dispatch methods including the scenario approach are introduced in Section III. Some numerical examples are given in Section IV to show the impact of DR uncertainty on day-ahead market, how scenario approach works, as well as ISO's preference among DRPs when congestion exists in the system. We conclude with our findings in Section V.

\section{Problem Formulation}

In this section the clearing process of day-ahead market in the presence of DRPs is illustrated. We briefly describe the strategy by which a DRP makes its offer, including price and ``capacity". The impact of a DRPs' uncertainty on the day-ahead scheduling is formulated, and an estimate of such uncertainty is made using empirical data obtained from the Electric Reliability Council of Texas (ERCOT).

\subsection{DRP as a Supplier in Day-ahead Market}
% part I: Introduction of Demand Response
In many organized regional markets, demand response (DR) is allowed to contribute to energy balance, capacity resource, and ancillary services \cite{Demand_History,CAISO_DR,CAISO_DR2,NYISO_DR,PJM_DR}. There are two types of DR: price-based and incentive-based. The main difference lies in the manner of encouraging end-consumers for their demand reduction: the former uses direct price signals (such as real-time or peak-valley prices), while the latter uses indirect or hybrid incentives \cite{US_DE}.

In this paper we concentrate on an ISO's day-ahead scheduling of energy resources including \emph{incentive-based DRPs}. DRPs aggregate DR commitments from end-consumers, and are treated as ``virtual generators" on the supply side in the day-ahead market. Compared to generators owned by generation companies (GenCos), a DRP's offer price is treated as its marginal cost, while the maximum DR commitment is treated as its ``capacity''.

\begin{figure}[!h]
\centering
\includegraphics[width=3.5in]{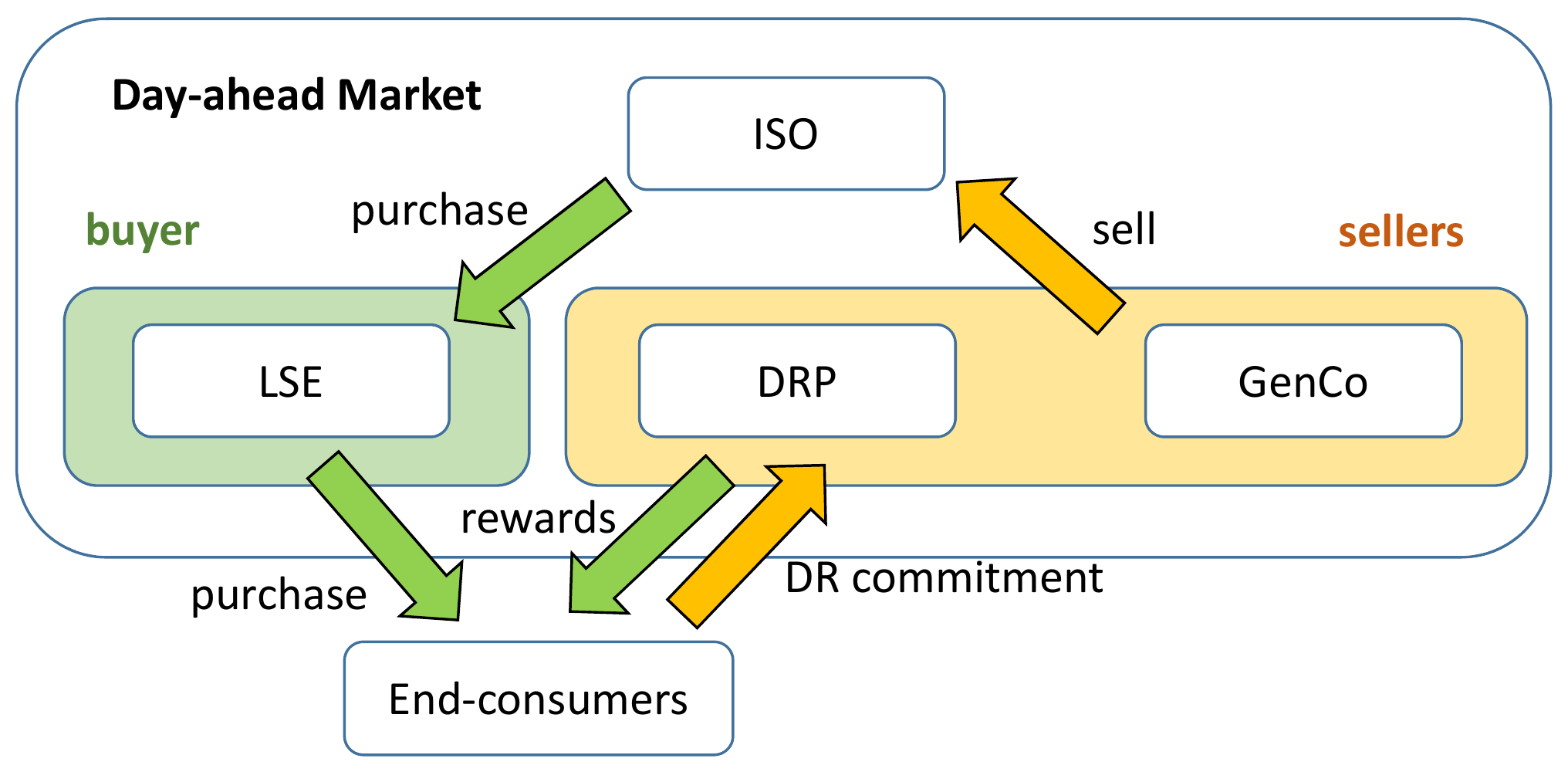}
\caption{Three-Layer Financial Structure of DAM with DRP}
\label{fig.DAM_Structure}
\end{figure}

The clearing process of the day-ahead market with DRPs is shown in Table \ref{table.DAM_DRP}, and a three-layer financial structure is visualized in Fig. \ref{fig.DAM_Structure}.

\begin{table}[!h]
%\tiny
\caption{\label{table.DAM_DRP}Clearing process of day-ahead market with DRPs}
\centering
\begin{tabular}
{|p{8cm}|} \hline
% {cp{2.5in}}
      (1) DRPs incentivize their end-consumers to reduce their electricity consumption for certain periods in the next day, and collect their DR commitments;\\
      (2) DRPs decide their offer sets ($\pi_{DR,j}$, $\overline{P_{DR,j}}$), GenCos prepare cost curves for generators, and LSEs predict their local demand;\\
      (3) The ISO clears the day-ahead market. Locational marginal prices (LMPs), accepted generations and DR commitments $P_{DR,j}$ are determined;\\
      (4) DRPs receive payments according to the \emph{realized} DR amount $P_{DR,j}^{real}$ multiplied by the LMP; GenCos are paid according to accepted generation multiplied by the LMP\\
      (5) End-consumers pay electricity bills to the LSE, and receive rewards from the DRP according to their power reduction in real-time.\\ \hline
\end{tabular}
\end{table}

\subsection{Offer Set for a DRP}
% part III: How DRP decide its bidding price & amount? (DRP's decision curve & demand curve for end-users)
Unlike conventional generators, a DRP's offer set ($\pi_{DR,j}$, $\overline{P_{DR,j}}$) cannot be fully determined by physical principles. A model is developed in this subsection to describe a DRP's decisions in steps (1), (2) in Table \ref{table.DAM_DRP}, as well as the inner relationship between the offer price $\pi_{DR,j}$ and ``capacity'' $\overline{P_{DR,j}}$ in a DRP's offer set.

\begin{figure}[!h]
\centering
\includegraphics[width=3.6in]{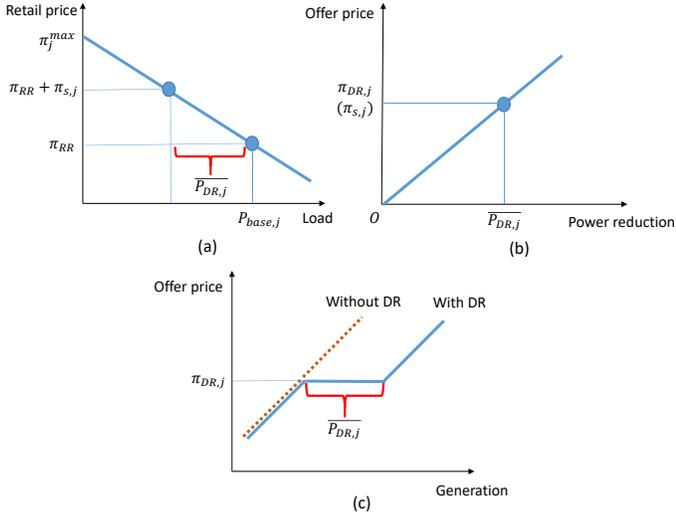}
\caption{DRP's Offering Strategy: (a) Inherent demand curve for end-consumers (b) DRP's offer curve (c) Market supply curve with the DRP}
\label{fig.DRP_Curves}
\end{figure}

Fig. \ref{fig.DRP_Curves}(a) shows the aggregated inherent demand curve for DRP $j$'s end-consumers. Load baseline $P_{base,j}$ is defined as the total electricity consumption of the end-consumers without any DR events; theoretically it equals the quantity on the demand curve corresponding to the retail price $\pi_{RR}$. It has been shown in \cite{CIDR_Hao} that, by claiming a certain reward $\pi_{s,j}$ per unit of power reduction, total electricity consumption shrinks as if the retail price increases by the same amount $\pi_{s,j}$. Therefore, the maximum DR commitment $\overline{P_{DR,j}}$ a DRP can collect equals the shrinkage of the load.

If a linear aggregated demand curve is assumed, $\overline{P_{DR,j}}$ can be calculated as
\begin{equation}\label{eq.PDRmax2}
\overline{P_{DR,j}}=min(P_{base,j},\frac{\pi_{s,j}}{\pi_{j}^{max}-\pi_{RR}}P_{base,j}),
\end{equation}
where $\pi_{j}^{max}$ is the Y-axis intercept of the demand curve.

Fig. \ref{fig.DRP_Curves}(a) is redrawn in \ref{fig.DRP_Curves}(b) to illustrate the relationship between the incentive price $\pi_{s,j}$ and the maximum DR commitment $\overline{P_{DR,j}}$. In practice, a DRP is able to choose any $\pi_{s,j}$ and collect the corresponding $\overline{P_{DR,j}}$ on the curve. Currently GenCos are required to submit the price-generation curve to the ISO \cite{ERCOT_truecost,CPUC_decision}. By assuming that DRPs also need to reveal their cost to the ISO, and that the incentive price is the only variable cost of DRPs, the offer price $\pi_{DR,j}$ should be equal to the incentive price $\pi_{s,j}$. By replacing $\pi_{s,j}$ with $\pi_{DR,j}$, Fig. \ref{fig.DRP_Curves}(b) represents the \emph{offer curve} of the DRP, which represents the relationship between its offer price and maximum DR commitment.

A DRP determines its offer by choosing an appropriate incentive price $\pi_{s,j}$ to its end-consumers. For simplicity, \emph{the strategy to choose the optimal $\pi_{s,j}$ is not discussed in this paper.} The maximum DR commitment $\overline{P_{DR,j}}$ is collected based on the inherent demand curve, and the DRP's offer price $\pi_{DR,j}$ is set equal to the incentive price. Unlike conventional generators which reveal their full cost curves to the ISO, only one point ($\pi_{DR,j}$, $\overline{P_{DR,j}}$) from the DRP's offer curve (\ref{fig.DRP_Curves}(b)) is submitted to the ISO during each day-ahead scheduling.

Fig. \ref{fig.DRP_Curves}(c) shows the change of the total supply curve after a DRP's offer is added into the day-ahead market. The total supply curve at the price $\pi_{DR,j}$ is shifted horizontally rightward by $\overline{P_{DR,j}}$. By choosing different points in its offer curve Fig. \ref{fig.DRP_Curves}(b) to formulate the offer set, a DRP can influence the shape of the supply curve in Fig. \ref{fig.DRP_Curves}(c), and therefore obtains different rewards in return after the market is cleared.

\subsection{Uncertainty of DR}
% part I: Why there is uncertainty in DR? (voluntary DR, and DR ratio is introduced)
In practice, a DRP's total demand reduction from the baseline usually deviates from its scheduled DR commitment $P_{DR,j}$ decided by the ISO in step (4), Table \ref{table.DAM_DRP}. Common factors that lead to the uncertainty of DR include consumers' uncertain behaviors, ambient condition change, load dynamics, etc. \cite{Uncertainty_Mathieu1,Uncertainty_Mathieu2}. The power imbalance or any constraint violation caused by such phenomena in real-time may result in additional re-dispatch cost to the system by means of ramping up expensive generators and the abandonment of renewable resources, etc. Therefore, it is crucial to find a suitable model to describe the uncertainty of DR.

To generalize the behavior of a group of consumers for a certain DRP $j$, the demand response (DR) ratio is defined as the ratio between the DRP's realized and scheduled DR amounts
\begin{equation}\label{eq.deltaDef2}
\delta_{j}=\frac{P_{DR,j}^{real}}{P_{DR,j}}, j=1,2,...,N.
\end{equation}

\begin{figure}[!h]
\centering
\includegraphics[width=2.4in]{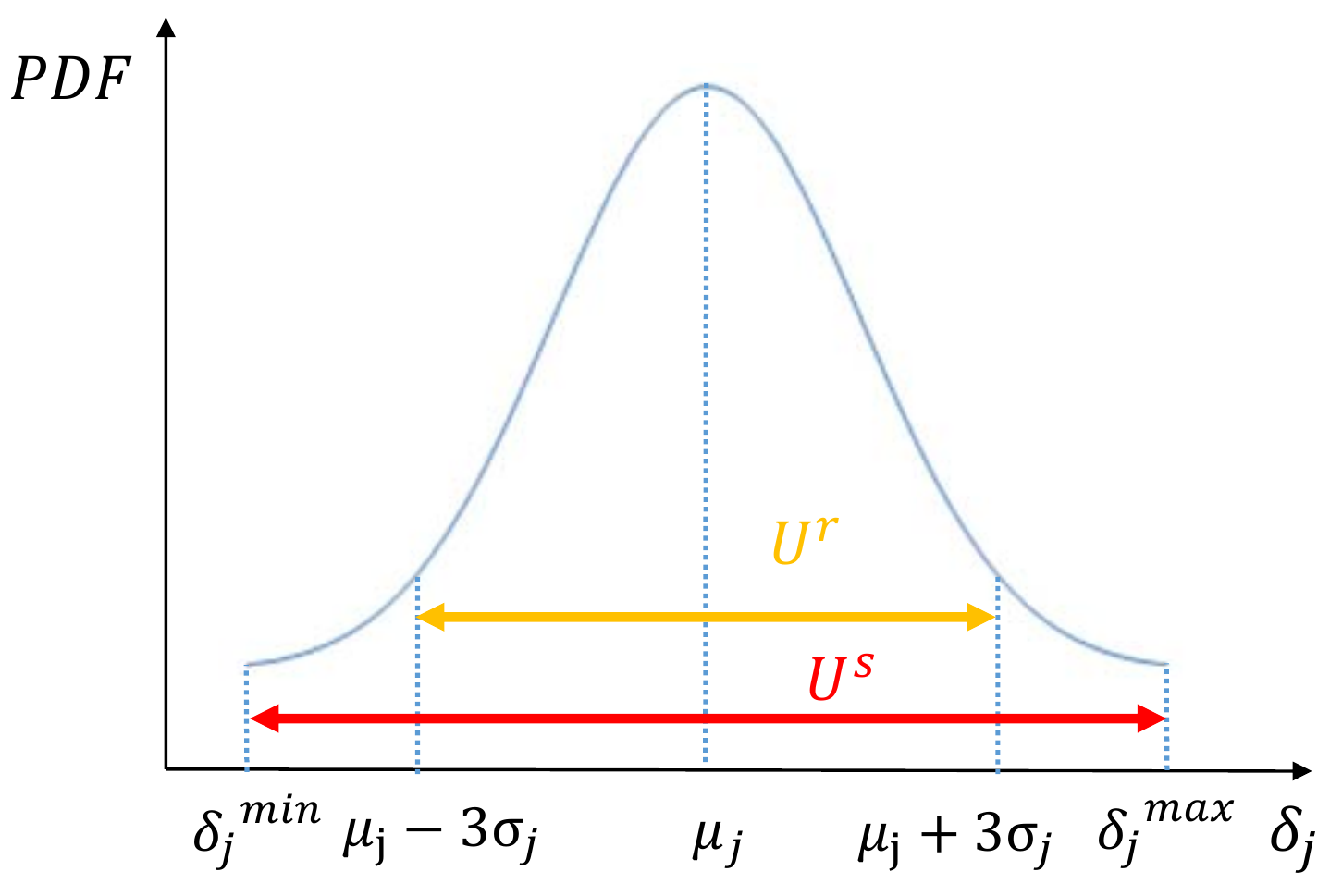}
\caption{Truncated Normal Distribution for Demand Response Ratio}
\label{fig.CutoffGaussian}
\end{figure}

In this paper, we make similar assumptions as \cite{Uncertainty_Mathieu1} that the realized DR amount $P_{DR,j}^{real}$ follows a \emph{truncated normal distribution}. Since $P_{DR,j}$ is a constant value submitted to the ISO, $\delta_{j}$ should also be a truncated normal distribution $\delta_{j} \sim N(\mu_{j},\sigma_{j}^{2},\delta_{j}^{min},\delta_{j}^{max})$ (Fig. \ref{fig.CutoffGaussian}).

Although no previous research has ever provided experimental data for incentive-based DRP, it has been shown in \cite{CIDR_Hao} that price-based and incentive-based DR are essentially equivalent; Therefore, price-based DR data of a commercial consumer in ERCOT area \cite{An_DR} is taken as a substitution. By assuming \emph{a virtual DRP} having only one such consumer ($j=1$), the uncertainty for this particular user can be calculated to be $\hat{\mu_{1}}=1$ and $\hat{\sigma_{1}}=0.67$.

The above assumptions may lead to an over-estimation of the uncertainty in incentive-based DRPs. First, consumer(s) using price-based DR are liable to become numb to too frequent changes of real-time prices, while incentive-based DR usually has less DR events and consumers are more focused on each DR event. In addition, the end-consumer in \cite{An_DR} is not assigned to any DRP; it means that the consumer cannot receive timely notifications from the DRP about the upcoming DR events. However, this estimation of $\delta_{j}$ is still valuable, and will be used in Section IV on a treatment group to show the impact of uncertainty levels of DR on market clearing results.

%\begin{itemize}
%  \item As organized entity, DRPs are able to inform and incentivize their end-consumers in DR programs;
%  \item Unlike real-time price-based DR, end-consumers focus on less frequent DR events provided by DRPs, and has less uncertainty in their response.
%\end{itemize}

\section{Economic Dispatch Methods in Day-ahead Market}

In this section multiple economic dispatch models are introduced to describe the clearing process of the day-ahead market. The ISO's goal is to maximize social welfare (or minimize total supply cost). Its decision variable $\xi=[P_{G},P_{DR}]$ consists of two components: accepted generation $P_{G}=[P_{G,i}], i=1,2,...,M$, and accepted DR commitment $P_{DR}=[P_{DR,j}], j=1,2,...,N$.

Recalling the definition of \emph{demand response ratio} in (\ref{eq.deltaDef2}), let the vector $\delta=[\delta_{1},\delta_{2},...,\delta_{N}]$ describe the uncertainty of all DRPs in the power system. As mentioned above, $\delta_{j}$ follows a truncated normal distribution $\delta_{j} \sim N(\mu_{j},\sigma_{j}^{2},\delta_{j}^{min},\delta_{j}^{max})$. If not specified differently, DR ratios for different DRPs are independent of each other.

%$\delta_{j}$ is a random variable defined as the ratio the realized DR commitment to the accepted/scheduled DR commitment by the ISO.

%More generally, in the system containing multiple DRPs, vector $\delta^{k}$ is used to show a certain scenario $k$. The $j^{th}$ element $\delta_{j}^{k}$ is one realization for DRP $j$; usually $\delta_{j}^{k}$ is assumed to be \emph{cut-off Gaussian distribution}, and is independent of each other.

\subsection{Deterministic Model}
The deterministic model widely used in today's day-ahead market neglects all the uncertainties in the power system. In our problem, this is equivalent to suppose that all realized DR commitments are equal to their scheduled amount ($\delta_{j}=1, j=1,2,...,N$). Therefore the deterministic optimization problem is described as follows:
\begin{eqnarray}\label{eq.DetModel1}\begin{split}
\min_{\xi} \ l(\xi)\\
s.t. \quad f_{1}(\xi)=0\\
\quad f_{2}(\xi) \le 0\\
\quad \underline{\xi} \le \xi \le \overline{\xi}.
\end{split}
\end{eqnarray}
The above $l(\xi)$ represents the \emph{total cost on supply side} including generators and DRPs. As a ``virtual generator'', the cost of a DRP is assumed to be the multiple of its offer price $\pi_{DR,j}$ and the accepted DR commitment $P_{DR,j}$, namely
\begin{eqnarray}\label{eq.DetModel2}
l(\xi):=\sum_{i=1}^{M}(a_{i}P_{G,i}^{2}+b_{i}P_{G,i})+\sum_{j=1}^{N}P_{DR,j}\pi_{DR,j}.
%+VOLL\cdot P_{v}
\end{eqnarray}

Equality constraint $f_{1}(\xi)=0$ instead represents the power balance requirement of supply and demand. To this purpose,
\begin{eqnarray}\label{eq.DetModel3}
f_{1}(\xi)=-(\sum_{i=1}^{M}P_{Gi}+\sum_{j=1}^{N}P_{DR,j})+P_{L}.
\end{eqnarray}
Eventually, $f_{2}(\xi) \le 0$ is meant to represent branch flow constraints, by defining
\begin{eqnarray}\label{eq.DetModel4}
f_{2}(\xi)=H \cdot P_{inj}(\xi)-F^{max},
\end{eqnarray}
where $P_{L}$ represents the total system load, $H$ is the distribution factor matrix, and $P_{inj}(\xi)$ is a vector indicating net branch flow injections on each bus.

In addition, decision variables $\xi=[P_{G} \quad P_{DR}]$ should satisfy their capacity limits $\underline{\xi} \le \xi \le \overline{\xi}$, which are equivalent to
\begin{eqnarray}\label{eq.DetModel5}\begin{split}
\underline{P_{G,i}} \le P_{G,i} \le \overline{P_{G,i}}, i=1,2,...,M\\
0 \le P_{DR,j} \le \overline{P_{DR,j}}, j=1,2,...,N,
\end{split}
\end{eqnarray}
%Note that $P_{DR,j}^{max}$ and $\pi_{DR,j}$ are pre-determined values.

\subsection{Stochastic Model}
The stochastic model focuses on minimizing the \emph{expectation} of supply-side cost in the power system considering the uncertainties involved. Particularly, this paper concentrates on the uncertainty of realized DR commitment $P_{DR,j}^{real}$ caused by random behaviors of end-consumers (measured by DR ratios). Therefore, uncertainty set $U^{s}$ in the stochastic model is defined, as the collection of $N_{k}$ independent and identically distributed (i.i.d.) extractions of $\delta$, where $\delta$ is the vector whose elements are the DR ratios from the DRPs ($\delta_{j}$). In practice, $U^{s}$ is created by directly using historical data. It can also be formulated by Monte Carlo sampling from the known probability distribution of $\delta$.

The stochastic optimization problem is described as follows:
\begin{eqnarray}\label{eq.StoModel1}\begin{split}
\min_{\xi} \ E[l(\xi,\delta)] \quad\\
s.t. \quad \quad \tilde{f_{1}}(\xi) \le 0 \quad \quad\\
\quad \quad \quad f_{2}(\xi,\delta) \le 0, \forall \delta \in U^{s}\\
\underline{\xi} \le \xi \le \overline{\xi}. \quad \quad{}
\end{split}
\end{eqnarray}

Compared to (\ref{eq.DetModel2}), the cost function $l(\xi,\delta)$ is influenced by the realized DR ratio. Therefore, the objective function in stochastic model is chosen as the average cost
\begin{eqnarray}\label{eq.StoModel3}
E[l(\xi,\delta)]=E[\sum_{i=1}^{M}(a_{i}P_{Gi}^{2}+b_{i}P_{Gi})+\sum_{j=1}^{N}\underbrace{\delta_{j}P_{DR,j}}_{P_{DR,j}^{real}}\pi_{DR,j}].
\end{eqnarray}

In addition, the uncertainty of the DRPs may also cause energy inadequacy to the system. Chance constraints are used to describe the requirement of maintaining energy adequacy with a probability higher than a certain level $(1-\gamma)$
\begin{eqnarray}\label{eq.StoModel4}
Pr(\sum_{i=1}^{M}P_{Gi}+\sum_{j=1}^{N}\delta_{j}P_{DR,j}>P_{L}) \ge 1-\gamma.
\end{eqnarray}

In order to solve the chance constrained optimization problem at low computational cost, in (\ref{eq.StoModel1}) the constraint (\ref{eq.StoModel4}) is updated by the following linear approximation, obtained from the semi-analytical method proposed in \cite{Chance-constraint_Hao}:
\begin{eqnarray}\label{eq.StoModel5}
\underbrace{-(\sum_{i=1}^{M}P_{Gi}+\sum_{j=1}^{N}g_{j}(\delta_{j},\gamma)P_{DR,j})+P_{L}}_{\tilde{f_{1}}(\xi)} \le 0
\end{eqnarray}
where $g_{j}:=\mu_{j}+\sigma_{j}\Phi^{-1}(1-\gamma)$ is always a constant value if the distribution of $\delta_{j}$ is known and $\gamma$ is pre-determined.

In addition, branch flow constraints are required to be strictly satisfied for all the scenarios of $\delta \in U^{s}$, leading to
\begin{eqnarray}\label{eq.StoModel6}
\underbrace{H \cdot P_{inj}(\xi,\delta)-F^{max}}_{f_{2}(\xi,\delta)} \le 0, \forall \delta \in U^{s}.
\end{eqnarray}

\subsection{Robust Model}
The following robust model is preferred when the ISO pays more attention to minimizing the supply-side cost for the \emph{worst case} of the uncertain parameter $\delta$ (\ref{eq.RobModel1}). In the robust model, energy adequacy and branch flow constraints need to be satisfied for all the scenarios in an uncertainty set $U^{r}$. The problem is described as
\begin{eqnarray}\label{eq.RobModel1}\begin{split}
\min_{\xi} \max_{\delta \in U^{r}} l(\xi,\delta) \quad \quad\\
s.t. \quad f_{1}(\xi,\delta) \le 0, \forall \delta \in U^{r}\\
\quad f_{2}(\xi,\delta) \le 0, \forall \delta \in U^{r}\\
\underline{\xi} \le \xi \le \overline{\xi}, \quad \quad \quad \quad
\end{split}
\end{eqnarray}
where the quadratic cost function $l(\xi,\delta)$ and branch flow constraint function $f_{2}(\xi,\delta)$ have the same form as in (\ref{eq.StoModel3}) and (\ref{eq.StoModel6}). Energy adequacy constraint is described as
\begin{eqnarray}\label{eq.RobModel3}
f_{1}(\xi,\delta)=-(\sum_{i=1}^{N}P_{Gi}+\sum_{j=1}^{M}\delta_{j}P_{DR,j})+P_{L} \le 0.
\end{eqnarray}

The uncertainty set $U^{r}$ is determined by the $3-\sigma$ rule, that is, $U^{r}$ is a box whose sides are intervals ($\delta_{j}^{min},\delta_{j}^{max}$) with $\delta_{j}^{min}=\mu_{j}-3\sigma_{j}$ and $\delta_{j}^{max}=\mu_{j}+3\sigma_{j}$.

\subsection{Scenario Approach Model} \label{sub:Scenario_Approach}

The scenario approach introduced by Campi and Garatti \cite{Campi_2009,Campi_2011} is a refinement of the convex robust model. By selecting a finite number of scenarios from the uncertainty set $U^{s}$, this approach improves dispatch performance while guaranteeing a quantifiable risk level. Similar to the robust model (\ref{eq.RobModel1}), the scenario approach in economic dispatch is described as follows:
\begin{eqnarray}\label{eq.SceModel1}\begin{split}
\min_{\xi} \max_{\delta \in U'} l(\xi,\delta) \quad \quad\\
s.t. \quad f_{1}(\xi,\delta) \le 0, \forall \delta \in U'\\
\quad f_{2}(\xi,\delta) \le 0, \forall \delta \in U'\\
\underline{\xi} \le \xi \le \overline{\xi}, \quad \quad \quad \quad
\end{split}
\end{eqnarray}
where $l(\xi,\delta), f_{1}(\xi,\delta)$ and $f_{2}(\xi,\delta)$ are as in (\ref{eq.RobModel1}).

The only difference lies in the uncertainty set $U'$. References \cite{Campi_2009,Campi_2011,Campi_2008Exact,Care_2014Fast} have proposed multiple approaches in formulating $U'$ in scenario approach; In this paper, the number of scenarios ($N_{k}$) in the original uncertainty set $U^{s}$ is assumed to be finite, which is reasonable especially when $U^{s}$ is obtained from historical data. $U'$ is formulated by removing a certain number of scenarios $p$ from $U^{s}$. It can be shown that formulation (\ref{eq.SceModel4}) is equivalent to (\ref{eq.SceModel1}) (see e.g. references \cite{Hformulation_Ben,Hformulation_Ghaoui,Campi_2009}). 
\begin{eqnarray}\label{eq.SceModel4}\begin{split}
\min_{\xi,h} \quad h \quad \quad \quad \quad\\
s.t. \quad l(\xi,\delta) \le h, \forall \delta \in U' \\
\quad f_{1}(\xi,\delta) \le 0, \forall \delta \in U'\\
\quad f_{2}(\xi,\delta) \le 0, \forall \delta \in U'\\
\underline{\xi} \le \xi \le \overline{\xi} \quad \quad \qquad
\end{split}
\end{eqnarray}
The reason for transforming (\ref{eq.SceModel1}) to (\ref{eq.SceModel4}) is to reformulate it so that it can be easily handled by standard optimization solvers. Moreover, this new formulation allows for easy removal of scenarios by simply removing the corresponding constraints.

As a brief explanation to (\ref{eq.SceModel4}), note that formulation (\ref{eq.SceModel1}) is equivalent to
\begin{eqnarray*}\label{eq.SceModel5}\begin{split}
\min_{\xi,h} \quad h \quad \quad \quad \quad\\
s.t. \quad \max_{\delta \in U'}l(\xi,\delta) \le h \qquad\\
\quad f_{1}(\xi,\delta) \le 0, \forall \delta \in U'\\
\quad f_{2}(\xi,\delta) \le 0, \forall \delta \in U'\\
\underline{\xi} \le \xi \le \overline{\xi}, \quad \quad \qquad
\end{split}
\end{eqnarray*}
since the choice of $\xi$ that allows one to reduce $h$ as much as possible is that returning the minimum of $\max_{\delta \in U'} l(\xi,\delta)$; so $\max_{\delta \in U'} l(\xi,\delta) \le h$ is clearly equivalent to $l(\xi,\delta) \le h, \forall \delta \in U'$.

It is perhaps worth noticing that in (\ref{eq.SceModel4}), by removing $p$ scenarios from the uncertainty set, a total number of $3p$ constraints are removed since there are three constraints $l(\xi,\delta) \le h$, $f_{1}(\xi,\delta) \le 0$ and $f_{2}(\xi,\delta) \le 0$ corresponding to each scenario $\delta$.

The main result associated to (\ref{eq.SceModel1}) is as follows. Let $\epsilon \in (0,1)$ be a ``violation'' parameter and $\beta \in (0,1)$ be a confidence parameter. Then, Theorem 2.1 of \cite{Campi_2011} claims that if the following inequality is satisfied
\begin{eqnarray}\label{eq.SceModel3}
{p+d-1\choose p}\sum_{i=0}^{p+d-1}{N_{k}\choose i}\epsilon^{i}(1-\epsilon)^{N_{k}-i} \le \beta
\end{eqnarray}
(being d the dimensionality of $\xi$), then with probability no more than $1-\beta$, the optimal solution of (\ref{eq.SceModel1}), say $\xi^{*}$ satisfies all the constraints in (\ref{eq.RobModel1}) but at most an $\epsilon$ fraction of violation ($f(\xi^{*},\delta)>0$ with probability at most $\epsilon$). 

The highlight of this theorem is its universality. The result holds as long as the objective function is convex, regardless of the interpretation of the objective function, constraints or probability distribution of uncertain variables. In addition, a careful choice of removing scenarios can potentially lead to a higher performance with a quantifiable risk level in the sense of probability of violation of constraints. In practice, the finite uncertainty set $U^{s}$ can be formulated using historical data.

Removing scenarios from the uncertainty set may help to improve the performance of the objective function. At the same time, it increases the system risk level, which is mainly reflected in three aspects:
\begin{enumerate}[(i)]
  \item \emph{Power inadequacy}: Violation of $f_{1}(\xi^{*},\delta) \le 0$ indicates energy inadequacy in certain scenarios $\delta \in U^{s}$.
  \item \emph{Branch flow violation}: Violation of $f_{2}(\xi^{*},\delta) \le 0$ means that the optimal solution $\xi^{*}$ causes overflow in certain scenarios $\delta \in U^{s}$.
  \item \emph{$h$-violation}: Violation of $l(\xi^{*},\delta) \le h^\ast$ means that higher supply cost (lower social welfare) is possible when applying the ``optimal'' solution $\xi^{*}$ in certain scenarios in $U^{s}$; $h^\ast$ is an over-optimistic estimation.
\end{enumerate}
The pleasant message converged by Theorem 2.1 of \cite{Campi_2011} is that the risk is quantifiable since one is $1-\beta$ confident that any violation happens with probability less than $\epsilon$, where $\epsilon$ is computed from (\ref{eq.SceModel3}).

% discussion about methods of eliminating scanrios
An important feature of Theorem 2.1 of \cite{Campi_2011} is that it holds true irrespective of the scenario removal algorithm, i.e. it holds true irrespective on the strategy through which scenario to remove are chosen. As an example, reference \cite{Campi_2011} has defined the ``optimal'' scenario removal algorithm as the algorithm that ``among all possible eliminations of p scenarios out of total N'' implement the one that gives the best improvement on the objective function. However, reference \cite{Campi_2011,Campi_2015} also explain that such an ``optimal'' algorithm can be computationally intractable. Then, one can resort to suboptimal algorithms to remove scenarios, such as the greedy algorithm, which removes scenarios sequentially; sometimes even the greedy algorithm is computationally unaffordable \cite{Campi_2011,Campi_2015}, and other algorithms must be used.

In this paper, the following two algorithms for removing scenarios in the economic dispatch with uncertain DRPs are proposed:
\begin{enumerate}[(1)]
\item \emph{``Min'' algorithm}. This algorithm is based on the fact that under-reaction of DR would cause energy inadequacy of the power system in real-time. Usually more expensive generators need to be ramped up by the system operator, which leads to a higher realization cost. Therefore, this algorithm assumes that scenarios with low total DR $min(\sum_{j=1}^{N}\delta_{j}^{real}P_{DR,j})$ are removed with high priority.
\item \emph{``Center'' algorithm}. This algorithm is based on the fact that DRPs' deviations from their scheduled amounts (expectation) may lead to branch flows violations. In scenarios when branch flows are violated, expensive auxiliary generators are used by the ISO to satisfy branch flow constraints. Therefore, this algorithm assumes scenarios with largest DR deviations $max(\sum_{j=1}^{N}|\delta_{j}^{real}-\mu_{j}|P_{DR,j})$ need to be removed with priority.
\end{enumerate}
These two suboptimal algorithms are tractable vis-a-vis computational cost since they require neither sequential solution nor extra optimization.

Since these two algorithms focus on different aspects of performance improvement, they will have impacts on different types of constraint violations (Type (i),(ii),(iii)). Further discussions on this topic are given in Section \ref{sub:3-bus_Figure}.

The abbreviation of scenario selection ``\emph{percentage/algorithm}'' is used in the remainder of this paper. For instance, $20\%/min$ represents the action of removing the $20\%$ smallest $\delta$'s following ``min'' algorithm from the uncertainty set.

%the system operator could prefer one to the other by its preference. For example, it may prefer ``Min'' algorithm if the system operator weights transmission limit violation more than DR acceptance ratio (h-violation). Similar results will be shown in the numerical example (Fig. \ref{fig.CaseFigure}).

\subsection{Realization Cost}

Uncertainty in the realization of DR may lead to multiple constraint violations in real-time, as we have discussed in Section \ref{sub:Scenario_Approach}. Therefore, in real-time dispatch (namely, the balancing market), ISO may need to ramp up/down some expensive reserves or auxiliary generators in order to satisfy the power balance as well as branch flow limits.

In our modeling of real-time balancing dispatch, auxiliary generators that have constant marginal cost are assumed to be on the buses that have DRPs. Therefore, each auxiliary generator is responsible for making up the deviation of the realized DR from its day-ahead scheduling level. Real-time balancing cost $W_{rtd}$ is computed as the auxiliary generators' marginal cost multiplied by the absolute value of ramp up/down power (which is equal to the deviation of DR at the same bus),
\begin{equation}\label{eq.RealModel1}
W_{rtd}(\xi^{*},\delta)=\sum_{j}^{N}\pi_{aux,j}|(\delta_{j}^{real}-\mu_{j})P_{DR,j}|,
\end{equation}
and the total realization cost is the summation of expected realized dispatch cost and the expectation of balancing cost
\begin{equation}\label{eq.RealModel2}
W_{real}=E_{\delta}[l(\xi^{*},\delta)+W_{rtd}(\xi^{*},\delta)].
\end{equation}
Here $\pi_{aux,j}$ is the marginal cost for auxiliary generator $j$, and $\xi^{*}$ is the optimal value of $\xi$ calculated by different models in Section III A-D.

\section{Numerical Examples}
Three power systems with DRPs are analyzed in this section in order to compare the performance of the scenario approach with other economic dispatch methods. In addition, the trade-off between performance and system risk level, as well as the impact of distribution of $\delta$ on social welfare is also discussed.

\subsection{3-bus System with one DRP}
% The structure and parameters of the 3-bus system is shown in Figure \ref{fig.CaseStudy}.
\begin{figure}[!h]
\centering
\includegraphics[width=3.5in]{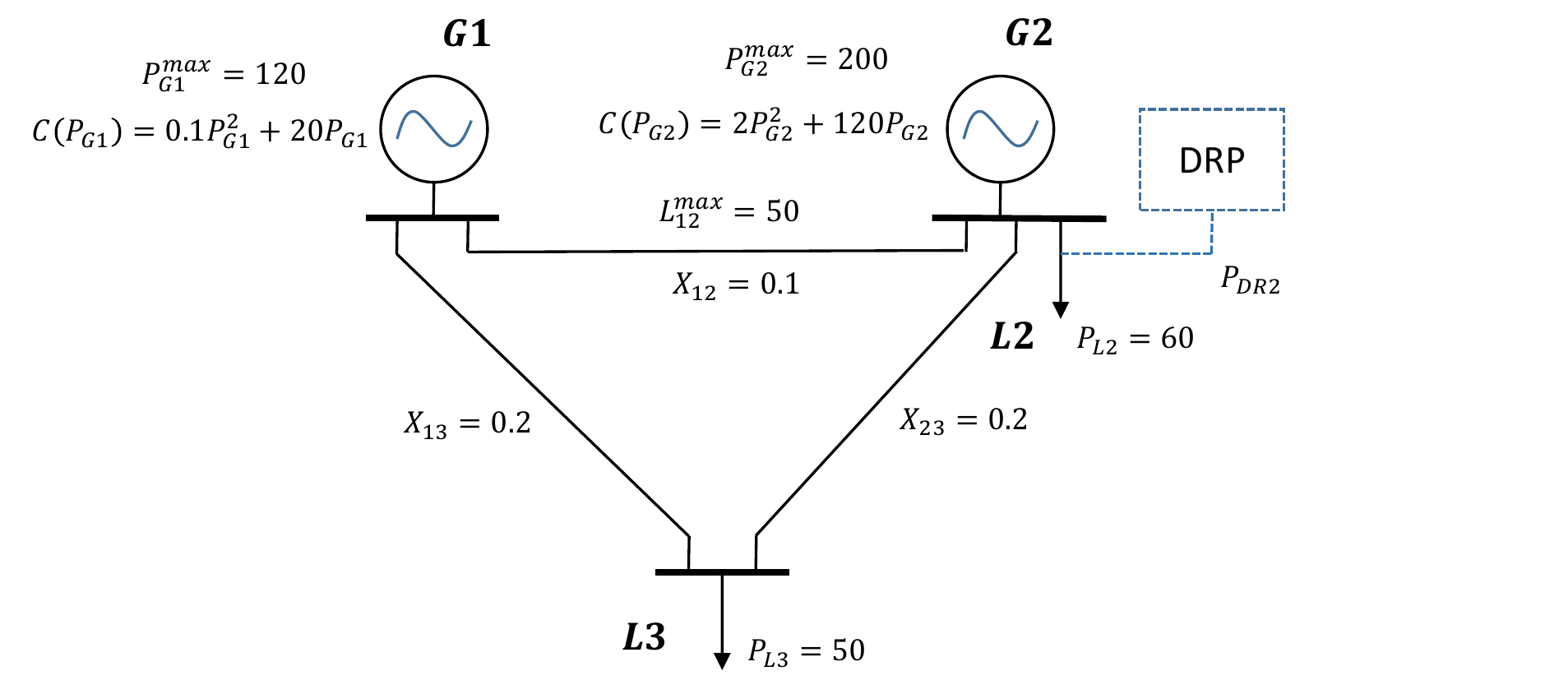}
\caption{3-bus System with the DRP}
\label{fig.CaseStudy}
\end{figure}

The 3-bus system shown in Fig. \ref{fig.CaseStudy} contains two generators, and one DRP on bus 2. \emph{All subscripts $j$ are dropped here since there is only one DRP}. Recalling Fig. \ref{fig.DRP_Curves}(a), a linear aggregated inherent demand curve for end-consumers is assumed, with baseline $P_{base}=P_{L2}=60$, retail price $\pi_{RR}=100$, and Y-axis intercept $\pi^{max}=400$.

Historical data of DR ratio  $U^{s}$ is created by randomly selecting $N_{k}=1000$ points from the truncated normal distribution ($\mu=1, \sigma=0.1, \delta^{max}=1.5$ and $\delta^{min}=0.5$). $U^{r}$ follows the $3-\sigma$ rule  ($\delta^{max}=0.7,\delta^{min}=1.3$). The reliability factor $\gamma=0.8$ in the stochastic model, $\beta=10^{-5}$ in the scenario approach, and $\pi_{aux,j}=150, \forall j$. The size of the uncertainty set is 1000.

\subsection{Simulation Results for Economic Dispatch} \label{sub:3-bus_Figure}
\begin{figure*}[!t]
\centering
\includegraphics[width=6in]{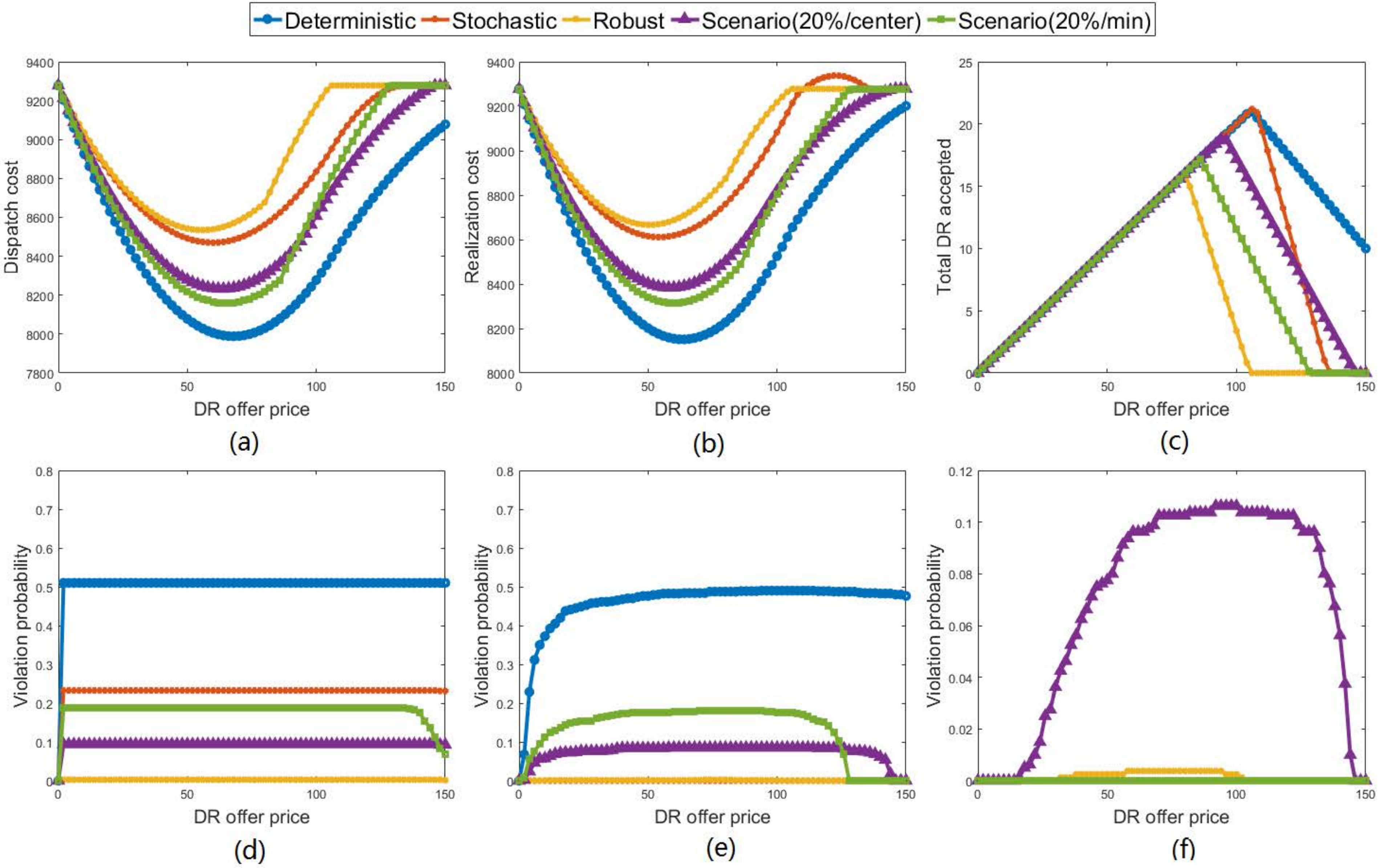}
\caption{Dispatch Results Under Different Models (a) Dispatch cost (b) Realization cost (c) DR bids accepted (d) Power inadequacy prob. (Type (i)) \protect\\ (e) Branch flow violation prob. (Type (ii)) (f) $h$-violation prob. (Type (iii))}
\label{fig.CaseFigure}
\end{figure*}

% part II: Economic dispatch result of different DR prices (figures and tables)
Simulation results for different economic dispatch models are shown in Fig. \ref{fig.CaseFigure}, and the results at one snapshot are shown in Table \ref{table.DispatchResult}. For all subgraphs in this figure, the $x$ axis represents the different choices of offer price $\pi_{DR}$ for the DRP. Higher offer price $\pi_{DR}$ results in more profit per unit DR, but the offer is less likely to be accepted by the ISO when $\pi_{DR}$ exceeds the marginal cost of other generations.

\emph{Economic dispatch performance is measured by the day-ahead dispatch cost and the realization cost, while system risk is measured by violation probability.} It can be concluded that the robust model is usually most conservative, since it takes the first inflection of DR acceptance in Fig. \ref{fig.CaseFigure}(c), and has the highest dispatch and realization cost in Fig. \ref{fig.CaseFigure}(a)(b). However, this level of conservatism leads to the lowest violation of all constraints (energy adequacy Fig. \ref{fig.CaseFigure}(d), branch flow limits Fig. \ref{fig.CaseFigure}(e) and $h$-violation status Fig. \ref{fig.CaseFigure}(f)). In contrast, deterministic model has the lowest dispatch and realization cost as well as the highest DR amount accepted. However, its neglect of uncertainty results in huge system risk in power adequacy and branch flow constraints. The stochastic model is relatively neutral in both performance and violation. Unlike the robust model, the deterministic and stochastic models don't measure the $h$-violation status since they purely concentrate on expected cost rather than cost in the worst case.

Fig. \ref{fig.CaseFigure} also shows the impact of scenario removal algorithms on different constraint violations (Type (i),(ii) and (iii)). Compared with a more \emph{neutral} ``center'' algorithm, the ``min'' algorithm is more \emph{optimistic} in the realization of DR, since it prefers removing low DR scenarios. The optimistic aspect of ``min'' algorithm is exhibited in Fig. \ref{fig.CaseFigure}(f) by low \emph{$h$-violation} probability on the dispatch cost. However, in realization, DR has higher probability to be lower than its expectation, which leads to \emph{lower DR acceptance ratio, and higher power inadequacy violation} (Fig. \ref{fig.CaseFigure}(d)). Given the fact than in our test system DR helps to relieve congestion, lower DR realization also leads to \emph{higher branch flow violation} (Fig. \ref{fig.CaseFigure}(e)).

The system operator may prefer to use the ``center'' over the ``min'' algorithm in this situation, since with similar levels of realization cost, the ``center'' algorithm provides higher DR acceptance ratio as well as lower physical violations (power adequacy and branch flows).

\subsection{Distributionally-Robust Feature of the Scenario Approach}

A significant advantage of the scenario approach is that the constraint violation upper bound is guaranteed by the theory (\ref{eq.SceModel3}). \emph{Therefore, optimization by the scenario approach doesn't rely on any assumption of the distribution of the underlying uncertainty.} In contrast, when using a stochastic model, such an assumption is necessary. Table \ref{table.DispatchResult} shows how an incorrect assumption on the distribution can affect the dispatch result. We take a snapshot at $\pi_{DR}=100$, and the case ``Sto(uniform)'' is created with the ``incorrect'' assumption that $\delta$ is uniformly distributed $\delta \sim U(0,2)$. Compared with the true model ``Sto'' ($\delta$ follows truncated normal distribution), ``Sto(uniform)'' clearly over-estimates the uncertainty by using the incorrect distribution; therefore, it schedules less efficient conventional generation in the day-ahead market, which leads to higher dispatch cost, realization cost and violation probability.

\begin{table*}[!t]
\caption{\label{table.DispatchResult}Dispatch Results under $\pi_{DR}=100$}
\centering
\begin{threeparttable}
\begin{tabular}
{c|cccccc}
% {cp{2.5in}}
      $Model \tnote{a}$ & Dispatch Cost & Realization Cost & Total DR & Balance Vio. & Branch Flow Vio. & $h$-Vio.\\ \hline
      $Dtm$ & $8278.1$ & $8520.9$ & $20.0$ & $0.496$ & $0.488$ & $N/A$\\
      $Sto$ & $8704.1$ & $8946.8$ & $20.0$ & $0.215$ & $0$ & $N/A$\\
      $Rob$ & $9127.9$ & $9181.1$ & $4.4$ & $0.0025$ & $0.0013$ & $0$\\
      $Sce(min)$ & $8577.5$ & $8792.8$ & $17.7$ & $0.104$ & $0.095$ & $0.105$\\
      $Sce(center)$ & $8556.2$ & $8727.3$ & $14.1$ & $0.210$ & $0.194$ & $0.013$\\ \hline
      $Sto(uniform)$ & $8718.0$ & $8963.3$ & $20$ & $0.232$ & $0$ & $N/A$\\ \hline
\end{tabular}
\begin{tablenotes}
  \item[a] Acronyms for models: Dtm: Deterministic, Sto: Stochastic, Rob: Robust, Sce: Scenario approach.
\end{tablenotes}
\end{threeparttable}
%\begin{tablenotes}
%  \item[1] *It can be proved that in LSE-base DR and DRP cases we can always get the same solution for $P_{GC}$, $P_{LC}$, $P_{DR}$ and $\pi_{DA}^{c}$.
%\end{tablenotes}
\end{table*}

% part III: Scenario approach: trade-off between realization cost & violation probability (figures included)
\subsection{Trade-off between Performance and Risk}

\begin{figure}[!h]
\centering
\includegraphics[width=3.6in]{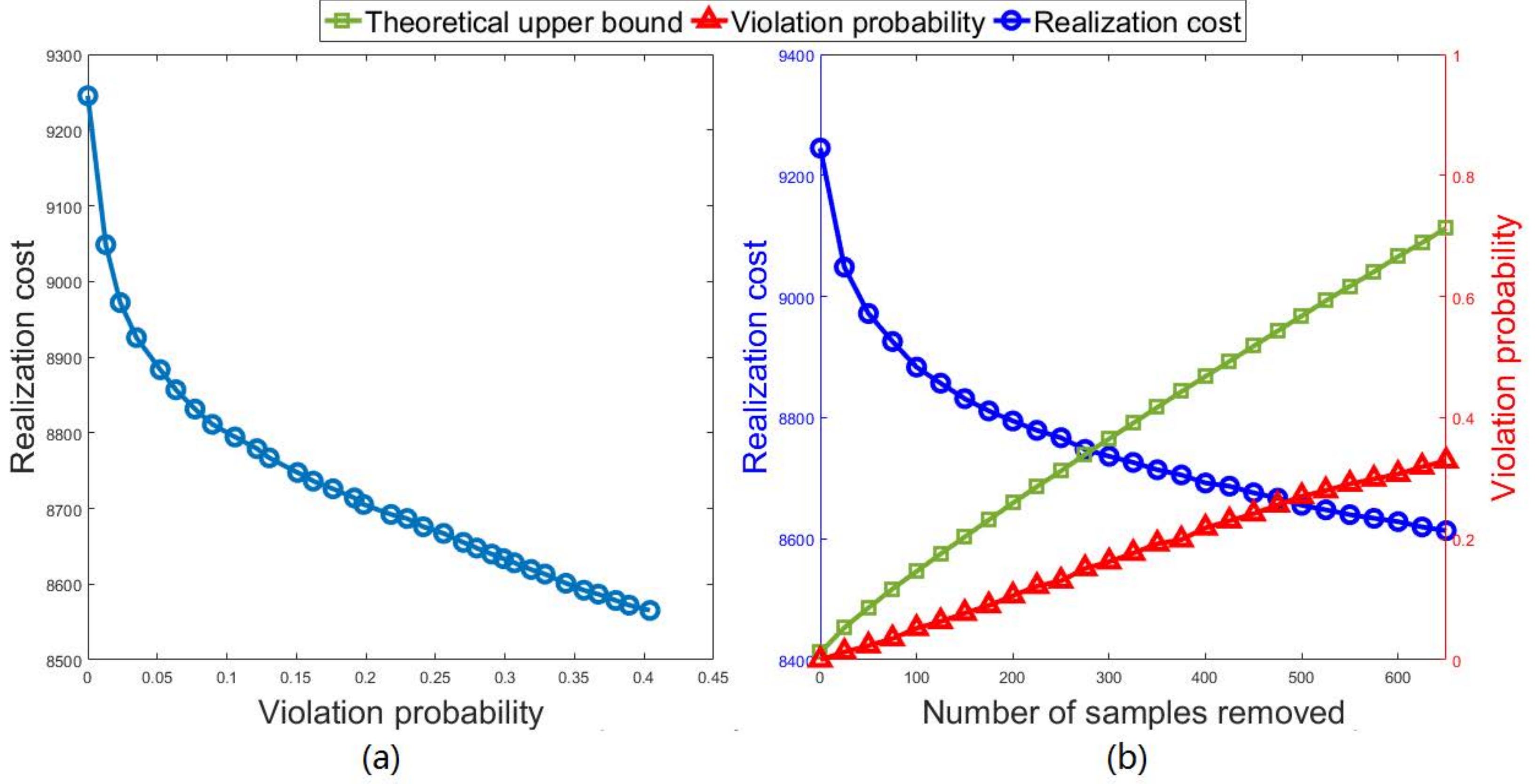}
\caption{Trade-off between cost and violation in scenario approach \protect\\ (a) Dispatch cost vs. violation probability (b) Dispatch cost and violation probability vs. number of removed samples}
\label{fig.CaseFigure2}
\end{figure}

As mentioned before, the scenario approach enables the trade-off between performance and system risk. Fig. \ref{fig.CaseFigure2} illustrates this trade-off by selecting the ``center'' algorithm to eliminate scenarios, and taking the snapshot of offer price $\pi_{DR}=100$.

Fig. \ref{fig.CaseFigure2}(a) directly shows the inverse proportional relationship between the realization cost and system risk. Fig. \ref{fig.CaseFigure2}(b) illustrates that with an increase of the number of scenarios $p$ removed, realization cost plunges immediately but becomes stable when $p>600$. This indicates a decreasing efficiency in improving the realization cost in scenario approach. On the other hand, the increase of violation probability is close to a linear increase and within the theoretical upper bound (\ref{eq.SceModel3}). The observations in (a)(b) could inspire the decision makers about how to choose an optimal $p$ in the scenario approach.

% part IV: Distribution changes & Error in estimation of distribution

\subsection{Influence of $\delta$ Distribution on DR acceptance}

\begin{figure}[!h]
\centering
\includegraphics[width=3.4in]{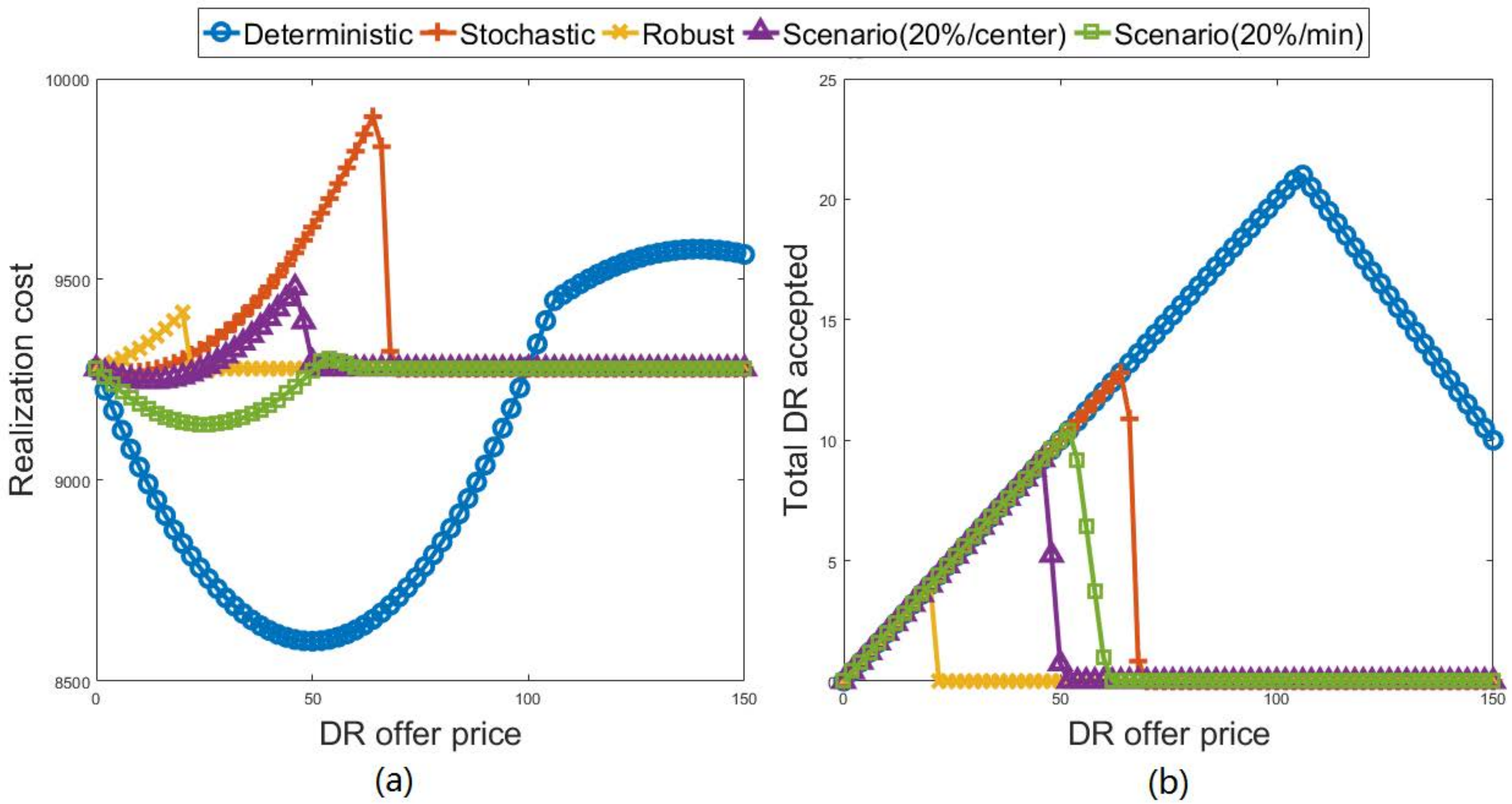}
\caption{Dispatch Result Under Different Models ($\sigma=0.67$) \protect\\ (a) Realization cost (b) DR amount accepted}
\label{fig.CaseFigure3}
\end{figure}

As discussed above, the distribution of $\delta$ has a significant impact on the realization cost as well as the system risk. Simulation results of higher uncertainty in the DRP ($\hat{\mu}=1, \hat{\sigma}=0.67, \delta^{max}=1.8$, $\delta^{min}=0.2$, calculated in Section II-C) on 3-bus system, are shown in Fig. \ref{fig.CaseFigure3}.
Compared to Fig. \ref{fig.CaseFigure}(a)(b), this figure shows a shrinking of DR acceptance and increase of total realization cost. Due to less reliability for the DRP, less DR commitment is taken by the ISO in order to maintain a required system risk level. High violation probability also causes increase in the realization cost.

\subsection{IEEE 14-bus System with 2 DRPs} \label{sub:IEEE14}

\begin{figure}[!h]
\centering
\includegraphics[width=2.8in]{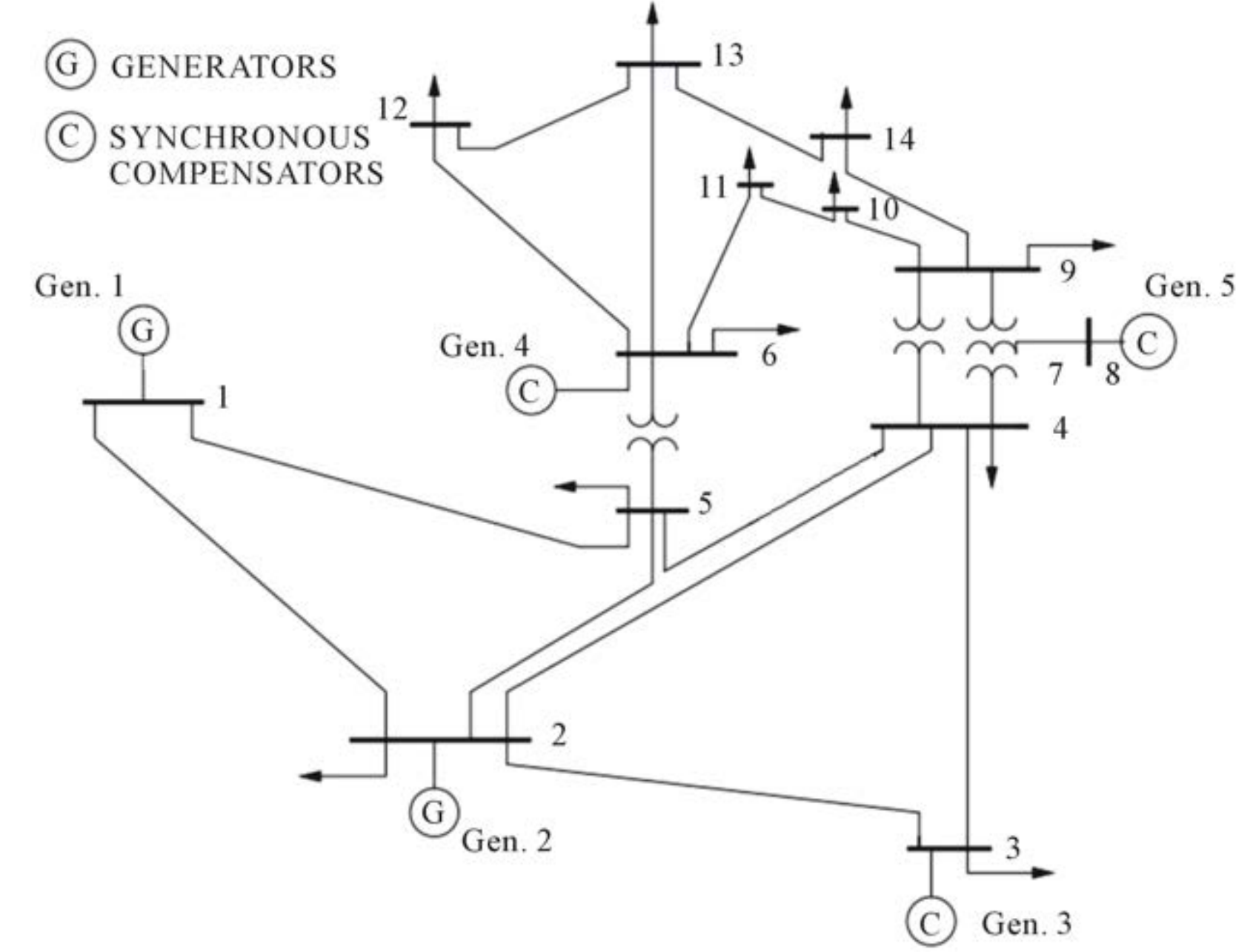}
\caption{IEEE 14-bus test system}
\label{fig.CaseFigure14bus4}
\end{figure}

In this subsection a more complicated IEEE 14-bus system with two DRPs is analyzed (Fig. \ref{fig.CaseFigure14bus4}). We assume that the two largest loads (buses 3 and 4) have DRPs ($N=2$) that can help to reduce the demand, with $\pi_{RR}=100,\pi_{c,j}^{max}=400, j=1,2$. Both DRPs have the same distribution of $\delta_{j}$ as that in Section IV-A, but are independent of each other. Line constraint $L_{24}=30$, reliability factor $\gamma=0.8$, and $\pi_{aux,j}=150, \forall j$. The size of the uncertainty set is 1000.

The ``center'' algorithm is adopted to select and remove scenarios in our simulation. However, the definition of this algorithm changes with multiple DRPs. Scenarios $\delta=[\delta_{1},\delta_{2}]$ whose DR commitment has most deviation from the expected total DR commitment,
\begin{equation}\label{eq.CenterDef}
\max_{\delta}(\sum_{j=1}^{N}|\delta_{j}-\mu_{j}|\overline{P_{DR,j}}),
\end{equation}
would be preferred to be removed. By definition, $P_{DR,j},j=1,2$ should be used in (\ref{eq.CenterDef}) in order to formulate the accepted DR commitment by the ISO. However, since scenarios are removed before the optimization problem is solved by the ISO, and the value of $P_{DR,j}$ is unknown, we use $\overline{P_{DR,j}},j=1,2$ instead as their rough estimates.

\begin{figure*}[!t]
\centering
\includegraphics[width=7in]{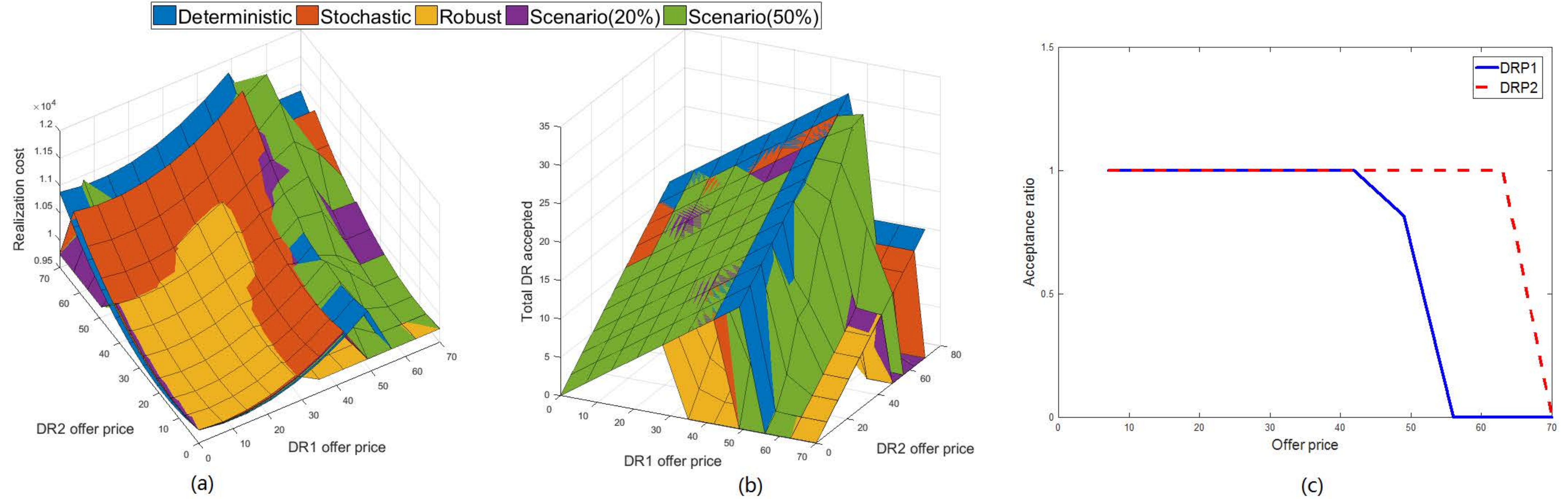}
\caption{IEEE 14-bus system with 2 DRPs (a) Realization cost (b) DR offers accepted (c) DR acceptance ratio ($\pi_{DR,1}=\pi_{DR,2}$)}
\label{fig.CaseFigure14bus1}
\end{figure*}

The simulation result in Fig. \ref{fig.CaseFigure14bus1}(a)(b) shows that the robust and stochastic models are more conservative than the scenario-approach, with earlier inflection of DR acceptance and higher realization cost. With an increasing number of scenarios removed, the scenario approach ($50\%$) has higher DR acceptance ratio than that of scenario approach ($20\%$).

% the following paragraph compare runtime for different methods
\begin{table}[!h]
%\tiny
\caption{\label{table.RunTime}Program computational time}
\centering
\begin{tabular}
{c|cc}
% {cp{2.5in}}
      Model & Mean Time(s) & Median Time(s)\\ \hline
      Dtm & $0.031$ & $0.026$\\
      Sto & $16.76$ & $13.61$\\
      Rob & $20.71$ & $17.14$\\
      Sce(20\%) & $14.04$ & $12.98$\\
      Sce(50\%) & $7.63$ & $5.84$\\ \hline
\end{tabular}
%\begin{tablenotes}
%  \item[1] *It can be proved that in LSE-base DR and DRP cases we can always get the same solution for $P_{GC}$, $P_{LC}$, $P_{DR}$ and $\pi_{DA}^{c}$.
%\end{tablenotes}
\end{table}
The computational cost is very similar among scenario approach (no scenario is removed), stochastic and robust models, since they have a similar number of decision variables as well as number of constraints. By removing scenarios from the uncertainty set, with a decrease in the number of constraints, the scenario approach runs faster than the other two approaches. Table III shows that in the 14-bus system, by removing 50\% of the scenarios, the computational time of the scenario approach is only 40\% of the stochastic model, and 35\% of the robust model.

Fig. \ref{fig.CaseFigure14bus1}(c) shows the influence of DRPs' locations on the accepted DR commitments in economic dispatch. Offer price $\pi_{DR,1}=\pi_{DR,2}$ is assumed in order to make the two DRPs comparable; the only difference lies in the truth that demand reduction on DRP 2 would help to release the congestion in the power system but DRP 1 would not. With the increase of offer price by both DRPs, the acceptance ratio (defined as $P_{DR,j}/\overline{P_{DR,j}}, j=1,2$) of DRP1 drops much earlier than that of DRP2. DRP2's offer is still accepted by the ISO at a price between $56$ and $70$, while the offer by DRP1 already gets rejected with the same offering price.

In summary, our analysis shows that the location of the DRP is essential to optimal dispatch results when there is line congestion. DRP offers which can help to ease the line congestion are preferred by the ISO, and have higher chance to get accepted.

\subsection{IEEE 118-bus System with 2 DRPs}

\begin{figure*}[!t]
\centering
\includegraphics[width=5.5in]{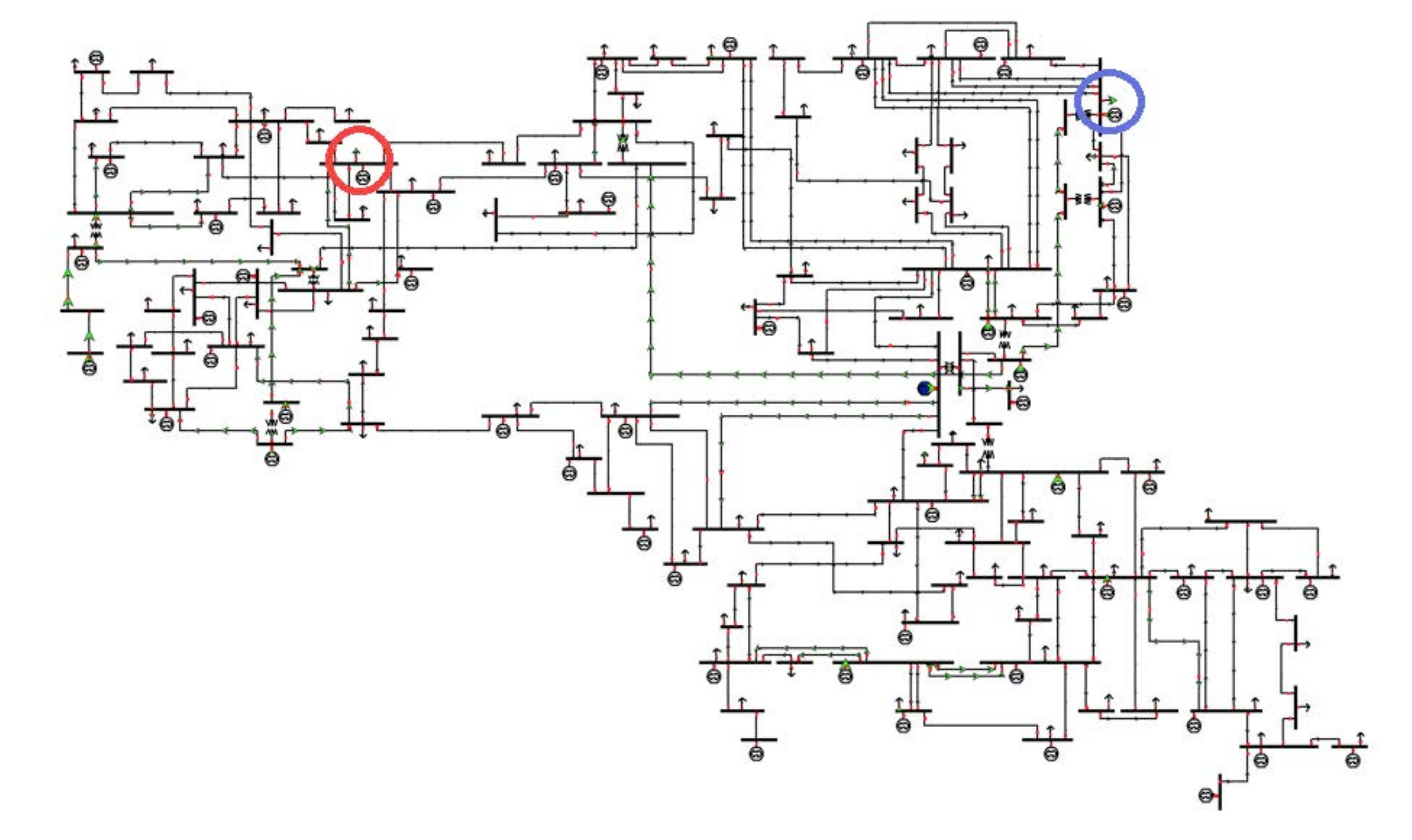}
\caption{IEEE 118-bus test system. Red marker: Bus 14, blue marker: Bus 59}
\label{fig.CaseFigure118}
\end{figure*}

In this subsection, a simulation is performed on an IEEE 118 bus system with two DRPs (Fig. \ref{fig.CaseFigure118}). We assume Buses 15 (markered with a red circle) and 59 (markered with a blue circle) have DRPs ($N=2$), with $\pi_{RR}=100,\pi_{c,j}^{max}=300, j=1,2$. There is no line flow limit in the system; $\gamma=0.8$, $\pi_{aux,j}=150, \forall j$, and the nuumber of scenairos in the uncertainty set $U^{s},U^{r},U^{test}$ are all $1600$.

In the simulation, the DRPs' offers are assumed to be $(\pi_{DR,1},\overline{P_{DR,1}})=(30,13.50)$ and $(\pi_{DR,2},\overline{P_{DR,2}})=(35,48.48)$. Simulation results are shown in Table \ref{table.IEEE118}. Some of the key observations include: (a) The robust-based dispatch exhibits the most conservative DR acceptance; (b) The scenario approach incurs less computational time than the stochastic and robust models; and (c), in terms of cost, it's not always better to remove more scenarios; in this case, too optimistic with DR in ``Sce(50\%)'' results in higher dispatch and realization cost.
\begin{table*}[!t]
%\tiny
\caption{\label{table.IEEE118}Simulation Results on IEEE 118-Bus System}
\centering
%\begin{threeparttable}
\begin{tabular}
{c|ccccccc}
% {cp{2.5in}}
      Model & Disp. Cost & Real. Cost & Total Gen & Total DR & Balance Vio. & h-Vio. & Computational Time (s)\\ \hline
      Dtm & $12562$ & $12812$ & $4180.0$ & $61.97$ & $0.516$ & - & $0.71$\\
      Sto & $12580$ & $12830$ & $4184.6$ & $61.97$ & $0.167$ & - & $6.58$\\
      Rob & $12595$ & $12595$ & $4242.0$ & $0$ & $0$ & $0$ & $2.42$\\
      Sce(20\%) & $12595$ & $12595$ & $4242.0$ & $0.34$ & $0.029$ & $0$ & $1.67$\\
      Sce(50\%) & $12812$ & $12814$ & $4188.0$ & $58.16$ & $0.173$ & $0.044$ & $1.52$\\ \hline
\end{tabular}
%\begin{tablenotes}
%  \item[a] Acronyms for models: Dtm: Deterministic, Sto: Stochastic, Rob: Robust, Sce: Scenario approach.
%\end{tablenotes}
%\end{threeparttable}
\end{table*}

% operational time: 0.12, 415.1, 31.79, 242.5, 95.1

\section{Concluding Remarks}

This paper introduces the new computational framework of scenario approach for accounting for the uncertainty of demand response providers (DRPs) in the day-ahead market. Unlike conventional generators, DRPs provide demand response by bottom-up aggregation, which causes high uncertainty of DR because of the behavior of end-consumers. The scenario approach model is introduced in order to obtain better performance than the robust model in most cases, while maintaining an acceptable system risk level. In addition, the scenario approach's key advantage lies in its ability to trade-off between the performance and system risk, compared with existing methods such as deterministic, robust and stochastic.

%slightly lower level of optimality in the worst scenario.
Simulation results shows that the uncertainty of DR will lead to less amount of DRPs' offer's being accepted in all dispatch methods (stochastic, robust and scenario approach). Though depending on the algorithm employed for selecting and removing scenarios, the scenario approach provides an improvement in expected realization cost over the robust model. In the economic dispatch, the decision maker is free to select and remove a number of scenarios from the uncertainty set in order to lower the realization cost. However, the removal of some scenarios comes at the cost of a small level of violation probability in the constraints. This paper also provides a preliminary assessment of scenario removal algorithms. This may help the system operator to increase the efficiency of improving the dispatch performance in the scenario approach.

Future work will investigate the trade-off of performance and risk involved with the scenario removal algorithms in economic dispatch in order to take care of both optimality and computational cost. Strategic behavior of multiple DRPs in the market will also be investigated in the context of scenario-based economic dispatch framework.

% Future work: cross-region free-rider effect of LSEs; definition of baseline and how it cause problem; gaming and cooperation between the LSE and DRPs.

\section*{Acknowledgement}

The authors would like to thank Mr. M. Sadegh Modarresi and Mr. Xinbo Geng of Texas A\&M University for comments and discussions that greatly improved the manuscript.

\renewcommand\refname{Reference}
\bibliographystyle{IEEEtran}
\bibliography{IEEEabrv,ScenarioEDref}

\begin{thebibliography}{10}
\providecommand{\url}[1]{#1}
\csname url@samestyle\endcsname
\providecommand{\newblock}{\relax}
\providecommand{\bibinfo}[2]{#2}
\providecommand{\BIBentrySTDinterwordspacing}{\spaceskip=0pt\relax}
\providecommand{\BIBentryALTinterwordstretchfactor}{4}
\providecommand{\BIBentryALTinterwordspacing}{\spaceskip=\fontdimen2\font plus
\BIBentryALTinterwordstretchfactor\fontdimen3\font minus
  \fontdimen4\font\relax}
\providecommand{\BIBforeignlanguage}[2]{{%
\expandafter\ifx\csname l@#1\endcsname\relax
\typeout{** WARNING: IEEEtran.bst: No hyphenation pattern has been}%
\typeout{** loaded for the language `#1'. Using the pattern for}%
\typeout{** the default language instead.}%
\else
\language=\csname l@#1\endcsname
\fi
#2}}
\providecommand{\BIBdecl}{\relax}
\BIBdecl

\bibitem{CAISO_DR}
{California ISO}, ``Demand response \& proxy demand resource-frequently asked
  questions.''

\bibitem{NYISO_DR}
{New York Independent System Operator (NYISO)}, ``Day-ahead demand response
  program manual.''

\bibitem{PJM_DR}
J.~McAnany, ``2014 demand response operations--market activity report: May
  2015.''

\bibitem{Uncertainty_Mathieu1}
J.~L. Mathieu, D.~S. Callaway, and S.~Kiliccote, ``Examining uncertainty in
  demand response baseline models and variability in automated responses to
  dynamic pricing,'' in \emph{Decision and Control and European Control
  Conference (CDC-ECC), 2011 50th IEEE Conference on}.\hskip 1em plus 0.5em
  minus 0.4em\relax IEEE, 2011, pp. 4332--4339.

\bibitem{Uncertainty_Mathieu2}
J.~L. Mathieu, M.~Gonzalez~Vaya, and G.~Andersson, ``Uncertainty in the
  flexibility of aggregations of demand response resources,'' in
  \emph{Industrial Electronics Society, IECON 2013-39th Annual Conference of
  the IEEE}.\hskip 1em plus 0.5em minus 0.4em\relax IEEE, 2013, pp. 8052--8057.

\bibitem{Stochastic_Classic}
J.~Hetzer, C.~Y. David, and K.~Bhattarai, ``An economic dispatch model
  incorporating wind power,'' \emph{IEEE Transactions on energy conversion},
  vol.~23, no.~2, pp. 603--611, 2008.

\bibitem{Stochastic_Zhao}
J.~Zhao, F.~Wen, Y.~Xue, Z.~Dong, and J.~Xin, ``Power system stochastic
  economic dispatch considering uncertain outputs from plug-in electric
  vehicles and wind generators,'' \emph{Dianli Xitong Zidonghua(Automation of
  Electric Power Systems)}, vol.~34, no.~20, pp. 22--29, 2010.

\bibitem{Stochastic_Dhillon}
J.~Dhillon, S.~Parti, and D.~Kothari, ``Stochastic economic emission load
  dispatch,'' \emph{Electric Power Systems Research}, vol.~26, no.~3, pp.
  179--186, 1993.

\bibitem{Anupam_2014}
A.~A. Thatte, X.~A. Sun, and L.~Xie, ``Robust optimization based economic
  dispatch for managing system ramp requirement,'' in \emph{System Sciences
  (HICSS), 2014 47th Hawaii International Conference on}.\hskip 1em plus 0.5em
  minus 0.4em\relax IEEE, 2014, pp. 2344--2352.

\bibitem{Sun_Robust}
{\'A}.~Lorca, X.~A. Sun, E.~Litvinov, and T.~Zheng, ``Multistage adaptive
  robust optimization for the unit commitment problem,'' \emph{Operations
  Research}, 2016.

\bibitem{Campi_2009}
M.~C. Campi, S.~Garatti, and M.~Prandini, ``The scenario approach for systems
  and control design,'' \emph{Annual Reviews in Control}, vol.~33, no.~2, pp.
  149--157, 2009.

\bibitem{Campi_2011}
M.~C. Campi and S.~Garatti, ``A sampling-and-discarding approach to
  chance-constrained optimization: feasibility and optimality,'' \emph{Journal
  of Optimization Theory and Applications}, vol. 148, no.~2, pp. 257--280,
  2011.

\bibitem{ProbN1_Vrakopoulou}
M.~Vrakopoulou, K.~Margellos, J.~Lygeros, and G.~Andersson, ``Probabilistic
  guarantees for the n-1 security of systems with wind power generation,'' in
  \emph{Reliability and Risk Evaluation of Wind Integrated Power
  Systems}.\hskip 1em plus 0.5em minus 0.4em\relax Springer, 2013, pp. 59--73.

\bibitem{Demand_History}
D.~Hurley, P.~Peterson, and M.~Whited, ``Demand response as a power system
  resource,'' \emph{Synapse Energy Economics Inc}, 2013.

\bibitem{CAISO_DR2}
{California ISO}, ``Initial comments of the california independent system
  operator corporation on the proposed decision adopting demand response
  activities and budgets for 2009 through 2011.''

\bibitem{US_DE}
{U.S.Department of Energy}, ``Benefit of demand response in electricity market
  and recommendations for achieving them.''

\bibitem{CIDR_Hao}
H.~Ming and L.~Xie, ``Analysis of coupon incentive-based demand response with
  bounded consumer rationality,'' in \emph{North American Power Symposium
  (NAPS), 2014}.\hskip 1em plus 0.5em minus 0.4em\relax IEEE, 2014, pp. 1--6.

\bibitem{ERCOT_truecost}
{Electric Reliability Council of Texas}, \emph{ERCOT's Verifiable Cost Manual},
  [online] Available: \url{http://www.ercot.com/}.

\bibitem{CPUC_decision}
{The California Public Utilities Commission}, \emph{{CPUC} decision No.
  91-06-022}, [online] Available: \url{http://www.cpuc.ca.gov/}.

\bibitem{An_DR}
J.~An, P.~Kumar, and L.~Xie, ``On transfer function modeling of price
  responsive demand: An empirical study,'' in \emph{Power \& Energy Society
  General Meeting, 2015 IEEE}.\hskip 1em plus 0.5em minus 0.4em\relax IEEE,
  2015, pp. 1--5.

\bibitem{Chance-constraint_Hao}
H.~Ming and L.~Xie, ``A semi-analytical method to solve chance constrained
  stochastic economic dispatch with demand response providers,''
  \emph{Intelligent Systems Application to Power Systems (ISAP) 2017}, 2017.

\bibitem{Campi_2008Exact}
M.~C. Campi and S.~Garatti, ``The exact feasibility of randomized solutions of
  uncertain convex programs,'' \emph{SIAM Journal on Optimization}, vol.~19,
  no.~3, pp. 1211--1230, 2008.

\bibitem{Care_2014Fast}
A.~Car{\`e}, S.~Garatti, and M.~C. Campi, ``Fast—fast algorithm for the
  scenario technique,'' \emph{Operations Research}, vol.~62, no.~3, pp.
  662--671, 2014.

\bibitem{Hformulation_Ben}
A.~Ben-Tal, L.~El~Ghaoui, and A.~Nemirovski, \emph{Robust optimization}.\hskip
  1em plus 0.5em minus 0.4em\relax Princeton University Press, 2009.

\bibitem{Hformulation_Ghaoui}
L.~El~Ghaoui and H.~Lebret, ``Robust solutions to least-squares problems with
  uncertain data,'' \emph{SIAM Journal on matrix analysis and applications},
  vol.~18, no.~4, pp. 1035--1064, 1997.

\bibitem{Campi_2015}
A.~Carè, S.~Garatti, and M.~C. Campi, ``Scenario min-max optimization and the
  risk of empirical costs,'' \emph{SIAM Journal on Optimization}, vol.~25,
  no.~4, pp. 2061--2080, 2015.

\end{thebibliography}

\begin{IEEEbiography}[{\includegraphics[width=1in,height=1.25in,clip,keepaspectratio]{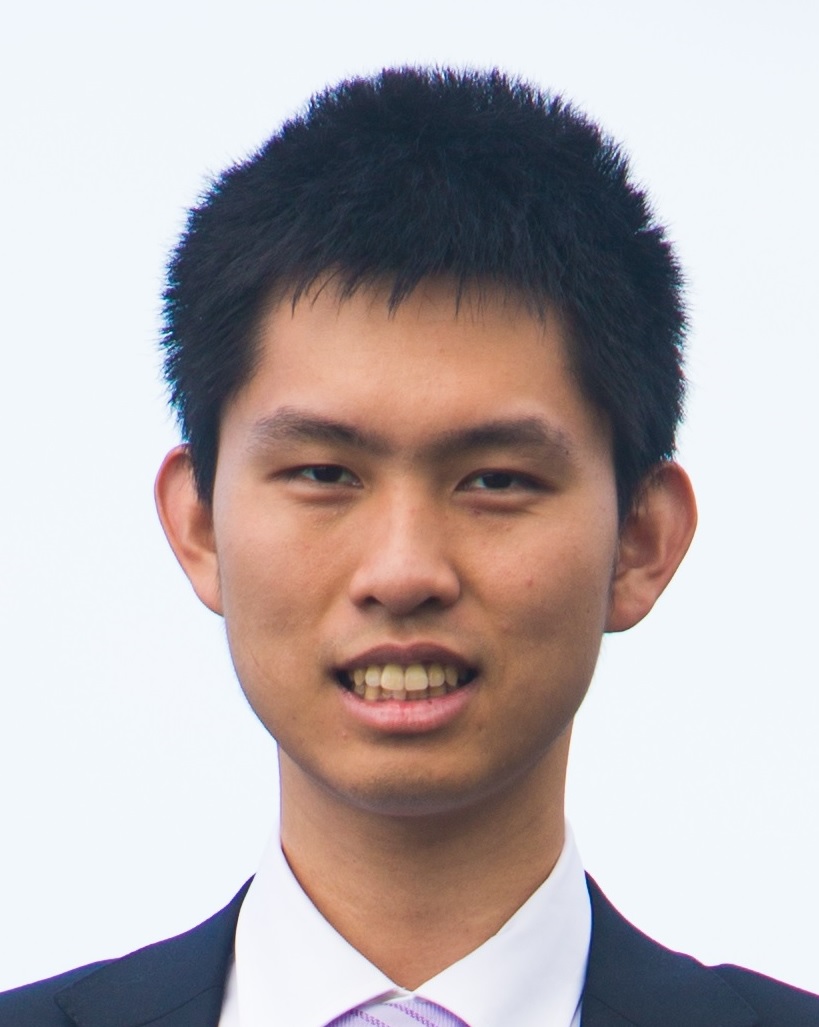}}]{Hao Ming}
(S'14) received his B.E. in Electrical Engineering and B.Econ in Economics from Tsinghua University, Beijing, China in 2012. He received M.S. in Electrical and Computer Engineering from Carnegie Mellon University in Dec 2013. He is pursuing his Ph.D degree in the department of Electrical and Computer Engineering at Texas A\&M University. His research interests include demand response, electricity markets, data analytics in power systems and low carbon electricity technology.
\end{IEEEbiography}

\begin{IEEEbiography}[{\includegraphics[width=1in,height=1.25in,clip,keepaspectratio]{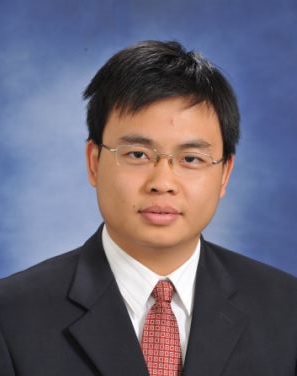}}]{Le Xie}
(M'10-SM'16) is an Associate Professor in the Department of Electrical and Computer Engineering at Texas A\&M University, where he is affiliated with the Electric Power and Power Electronics Group. He received his B.E. in Electrical Engineering from Tsinghua University, Beijing, China in 2004. He received S.M. in Engineering Sciences from Harvard University in June 2005. He obtained his Ph.D. from Electric Energy Systems Group (EESG) in the Department of Electrical and Computer Engineering at Carnegie Mellon University in 2009. His research interest includes modeling and control of large-scale complex systems, smart grid applications in support of renewable energy integration, and electricity markets.
\end{IEEEbiography}

\begin{IEEEbiography}[{\includegraphics[width=1in,height=1.25in,clip,keepaspectratio]{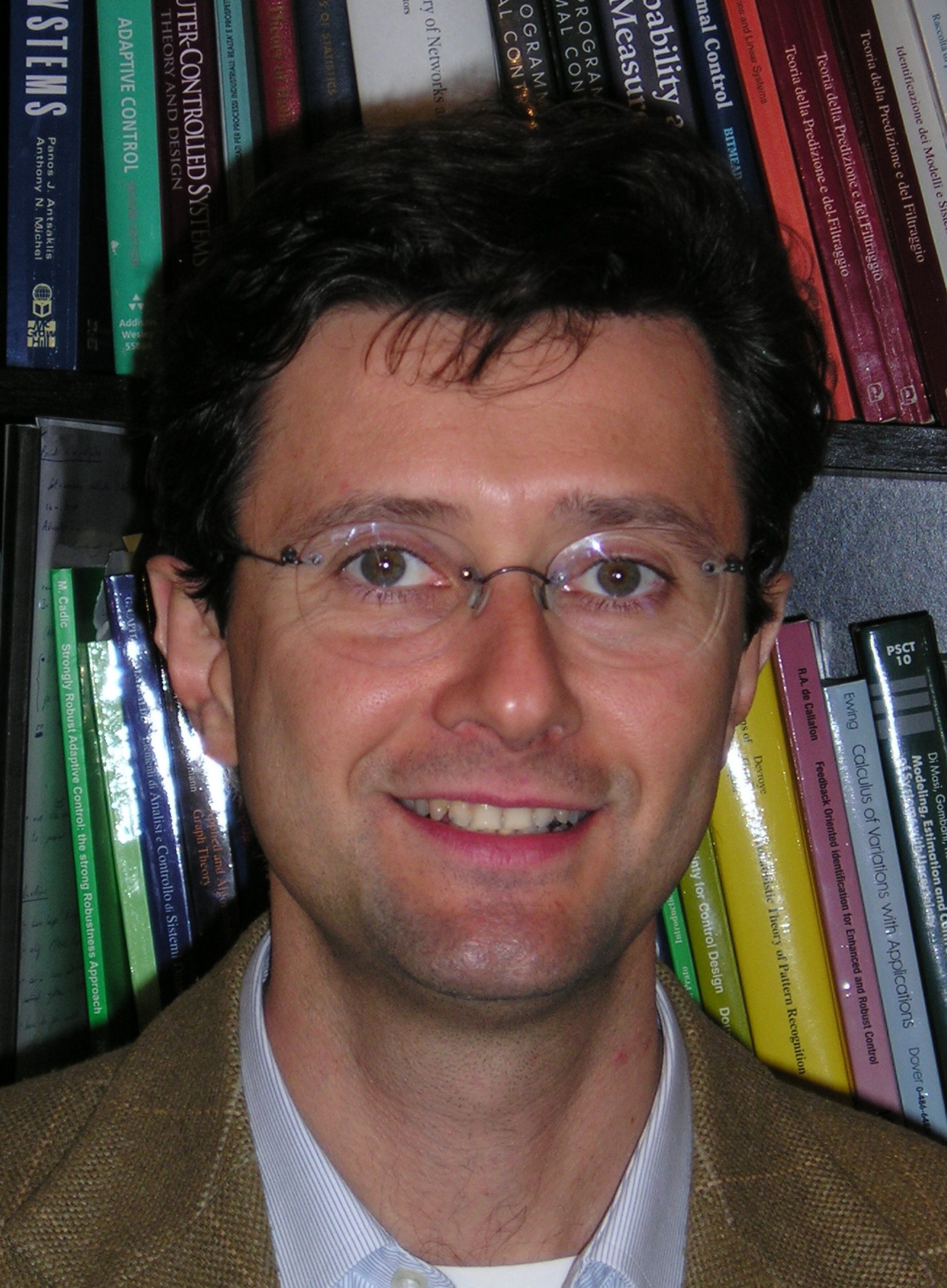}}]{Marco Claudio Campi}
(F'12) is professor of control at the University of Brescia, Italy. He is the chair of the Technical Committee IFAC on Modeling, Identification and Signal Processing (MISP) and has been in various capacities on the Editorial Board of Automatica, Systems and Control Letters and the European Journal of Control. Marco Campi is a recipient of the ``Giorgio Quazza'' prize, and, in 2008, he received the IEEE CSS George S. Axelby outstanding paper award for the article ``The Scenario Approach to Robust Control Design''. He has delivered plenary and semi-plenary addresses at major conferences including SYSID, MTNS, and CDC. Currently he is a distinguished lecturer of the Control Systems Society. Marco Campi is a Fellow of IEEE, a member of IFAC, and a member of SIDRA.
\end{IEEEbiography}

\begin{IEEEbiography}[{\includegraphics[width=1in,height=1.25in,clip,keepaspectratio]{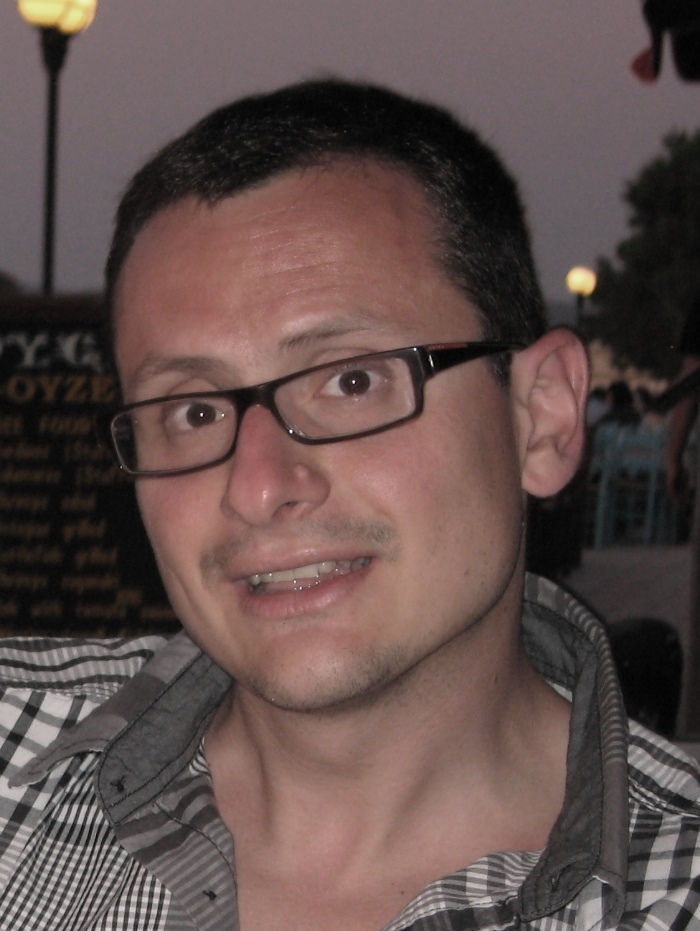}}]{Simone Garatti}
(M'12) is Associate Professor at the Dipartimento di Elettronica ed Informazione of the Politecnico di Milano, Milan, Italy. He received the Laurea degree and the Ph.D. in Information Technology Engineering in 2000 and 2004, respectively, both from the Politecnico di Milano. From 2005 to 2015 he was assistant professor at the same university. In 2003, he was visiting scholar at the Lund University of Technology, Lund, Sweden, in 2006 at the University of California San Diego (UCSD), San Diego, CA, USA, and in 2007 at the Massachusetts Institute of Technology and the Northeastern University, Boston, MA, USA.  He is member of  the IEEE Technical Committee on Computational Aspects of Control System Design and of the IFAC Technical Committee on Modeling, Identification and Signal Processing. His research interests include data-based and stochastic optimization for problems in systems and control, system identification, model quality assessment, and uncertainty quantification.
\end{IEEEbiography}

\begin{IEEEbiography}[{\includegraphics[width=1in,height=1.25in,clip,keepaspectratio]{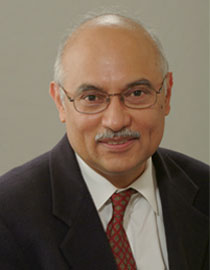}}]{P. R. Kumar}
(F'88) B. Tech. (IIT Madras, `73), D.Sc. (Washington University, St. Louis, `77), was a faculty member at UMBC (1977-84) and Univ. of Illinois, Urbana-Champaign (1985-2011). He is currently at Texas A\&M University. His current research is focused on cybersecurity, privacy, unmanned air systems, cyberphysical systems, energy systems, wireless networks, and automated transportation. 

He is a member of the US National Academy of Engineering, The World Academy of Sciences, and the Indian National Academy of Engineering. He was awarded a Doctor Honoris Causa by ETH, Zurich. He has received the IEEE Field Award for Control Systems, the Donald P. Eckman Award of the AACC, Fred W. Ellersick Prize of the IEEE Communications Society, the Outstanding Contribution Award of ACM SIGMOBILE, the Infocom Achievement Award, and the SIGMOBILE Test-of-Time Paper Award. He is a Fellow of IEEE and ACM Fellow. He was Leader of the Guest Chair Professor Group on Wireless Communication and Networking at Tsinghua University, is a D. J. Gandhi Distinguished Visiting Professor at IIT Bombay, and an Honorary Professor at IIT Hyderabad. He was awarded the Distinguished Alumnus Award from IIT Madras, the Alumni Achievement Award from Washington Univ., and the Daniel Drucker Eminent Faculty Award from the College of Engineering at the Univ. of Illinois.

\end{IEEEbiography}

\end{document}